\documentclass[aps,prd,groupedaddress,nofootinbib,superscriptaddress,preprint,11pt,a4paper]{revtex4-1}
\usepackage[colorlinks=true,citecolor=blue,linkcolor=blue,
breaklinks=true]{hyperref}
\usepackage{amsmath,amssymb}
\usepackage{graphicx,graphics}
\usepackage{array}
\usepackage{mathrsfs}
\usepackage{slashed}
\usepackage{float}
\usepackage{multirow}
\usepackage{booktabs}

\newcommand{\be}{\begin{equation}}
\newcommand{\ee}{\end{equation}}
\newcommand{\bea}{\begin{eqnarray}}
\newcommand{\eea}{\end{eqnarray}}

\begin{document}
\title{Jet Substructure Analysis for Distinguishing Left- and Right-Handed Couplings of Heavy Neutrino in $W'$ Decay at the HL-LHC}

\author{Songshaptak De}
\email{songshaptak.d@iopb.res.in}
\email{deswaptak@gmail.com}
\affiliation{Institute of Physics, Bhubaneswar, Sachivalaya Marg, Sainik School, Bhubaneswar 751005, India.}
\affiliation{Homi Bhabha National Institute, Training School Complex, Anushakti Nagar, Mumbai 400094, India.}

\author{Atri Dey}
\email{atri@stp.dias.ie}
\email{atridey1993@gmail.com}
\affiliation{School of Theoretical Physics, Dublin Institute for Advanced Studies,\\ 10 Burlington Road, Dublin 4, Ireland.}

\author{Tousik Samui}
\email{tousiks@imsc.res.in}
\email{tousiksamui@gmail.com}
\affiliation{The Institute of Mathematical Sciences, IV Cross Road, CIT Campus, Taramani, Chennai 600113, India.}
\preprint{IOP/BBSR/2024-09}
\preprint{DIAS-STP-24-21}
\preprint{IMSc/2024/03}

\begin{abstract}
The search for heavy $W'$ bosons in their decay modes to a lepton and a heavy neutrino offers a promising avenue for probing new physics beyond the Standard Model. This work focuses on such a signature with an energetic lepton plus a fat jet, originating from the heavy neutrino and containing a lepton. We have employed the jet substructure techniques to isolate the embedded lepton as a subjet of the fat jet. The Lepton Subjet Fraction ($LSF$) and Lepton Mass Drop ($LMD$) variables constructed from the lepton subjet help in separating the signal region from the background. We further study the polarization properties of the $W'$ coupling to the lepton and heavy neutrino through the decay products of the neutrino. Instead of relying on a specific model, we employ generic couplings and explore the discrimination power. Jet substructure-based angular variables $z_\ell$, $z_\theta$, and $z_k$ are combined to form BDT scores to obtain a better separation power between the left-chiral ($V-A$) and right-chiral ($V+A$) coupling configurations. By using $CL_s$ type profile likelihood estimator, we could achieve 1.6$\sigma$ -- 2.8$\sigma$ significance of excluding one coupling configuration in favour of the other.

\end{abstract}

\maketitle
\section{Introduction}
The triumph of the Standard Model (SM) of particle physics was marked by the observation of a 125-GeV scalar particle, closely resembling the SM Higgs boson, by CMS and ATLAS collaborations at the LHC\,\cite{CMS:2012qbp,ATLAS:2012yve}. Therefore, there is no denying the success of the SM, which has become the cornerstone of our understanding of fundamental interactions between elementary particles seen in nature. 
Despite its success, several theoretical and experimental indications suggest the need for new physics beyond the Standard Model (BSM).
From the theoretical perspective, we do not have a full understanding of the hierarchy of the masses of various flavours of the SM fermions, nor do we understand the lightness of the Higgs boson. These are referred to as the flavour hierarchy\,\cite{Froggatt:1978nt} and naturalness\,\cite{MORRISSEY20121,Farina:2013mla,deGouvea:2014xba} problems of the SM, respectively. More pressing, from the experimental side, the SM does not have any explanation for the masses and mixing of the three neutrino flavours\,\cite{K2K:2002icj,PhysRevLett.90.021802,RevModPhys.88.030501, RevModPhys.88.030502,Esteban:2020cvm,deSalas:2020pgw}. At the same time, on the cosmological and astrophysical side, there is no suitable explanation for the baryon asymmetry of the universe\,\cite{Sakharov:1967dj} nor for the existence of dark matter\,\cite{Arbey:2021gdg}. These unresolved issues compel us to venture beyond the SM to address some or all of these phenomena.

Assuming that the new physics beyond the SM exists, the SM can be thought of as an effective model, describing the particle interactions at the TeV scale, of an ultraviolet (UV)-complete model. In such UV-complete models, a natural consequence is the presence of one or more heavy particles. The models, which extend the SM gauge group, naturally predict additional vector bosons. On the other hand, the extension of the neutrino sector to address the neutrino mass and mixing puzzle opens up the possibility of heavier neutrinos. Such heavy particles can also arise without extending the gauge sector or particle sector in the SM. For example, the models with extra spatial dimensions, such as the Randall-Sundram (RS) model\,\cite{Randall:1999ee,Randall:1999vf}, predict heavier gauge bosons and heavier neutrinos through the Kaluza-Klein modes of the SM fields as a result of the compactification of the extra dimensions.

One of the compelling extensions to the Standard Model (SM) involves the introduction of additional heavy-charged vector bosons, commonly referred to as $W'$ bosons.
These hypothetical particles are thought to be a counterpart of the SM $W$ boson, which mediates the weak force.
Such a scenario arises in Left-Right Symmetric model (LRSM)\,\cite{Mohapatra:1974gc,Mohapatra:1974hk,Senjanovic:1975rk,Senjanovic:1978ev,ThomasArun:2021rwf}, where there are two $SU(2)$ gauge groups, one responsible for the SM particle interaction and the other dominantly couples to the right-handed fermions. A set of heavier neutrinos and gauge bosons are hence predicted, and coupling between $W'$ to SM or BSM particles is generated through mixing\,\cite{Han:2012vk,Cai:2017mow}. Heavy gauge bosons also naturally arise in models with extra dimensions, such as the Randall-Sundram (RS) model. These heavier modes of the SM fields arise as Kaluza-Klein excitations due to the compactification of the extra spatial dimensions. Notably, in the LRSM case, the $W'$ coupling to a pair of fermions is expected to be right-chiral in nature, while the coupling is expected to be left-chiral in the models with extra dimensions. One of our queries is whether or not we can distinguish if such a discovery is attained at the high-luminosity LHC (HL-LHC) at 14 TeV centre-of-mass energy.

Our method, in this work, does not consider a specific model, which usually considers either left-handed or right-handed coupling of $W'$. Rather, we take generic coupling of $W'$ to $q\bar{q'}$ and $\ell^\pm N$ ($N$ being neutrino or anti-neutrino). We then find signatures of $W'$ boson in the lepton plus a fat jet channel, where the fat jet comes from the decay of heavy neutrino. This fat jet also contains a lepton inside it. With the help of advanced jet substructure techniques and variables, we plan to isolate the embedded lepton as a subjet inside the fat jet. The Lepton Subjet Fraction ($LSF$) and Lepton Mass Drop ($LMD$) variables\,\cite{Brust:2014gia} are suggested to isolate such a lepton-subjet and would help us reduce the huge QCD background to a great extent. At the same time, we can discriminate between left- and right-handed coupling hypotheses considering $W'$ boson is found in the mentioned channel. The unique points of our study can be summed up to the following points:
\begin{itemize}
\item Our study focuses on the searches of heavy $W'$ signal in the channel with an energetic lepton and a fat jet, which embeds a lepton as a subjet.
\item Jet Substructure technique is employed to find this leptonic subjet and to help reduce background to a significant extent.
\item Polarization sensitive jet substructure observables are used to find the two different hypotheses of left- and right-handed coupling of $W'$ to the lepton and heavy neutrino pair.
\end{itemize}

We further note that the search for $W'$ bosons remains one of the major focuses of the current run of the LHC. The experimental searches typically involve looking for high-mass resonances in events with leptons and missing energy, which could indicate the production and decay of a $W'$ boson. The absence of discovery sets lower bounds on the mass of $W'$ bosons, pushing the limits into the multi-TeV range.  The current bound of $W'$ mass stands at somewhere around 5.5 TeV in the context of a sequential SM (SSM) scenario, where an SM-like coupling is considered for the $W'$\,\cite{CMS:2021dzb,ATLAS:2019fgd}. In particular, the searches in the channel $\ell\ell j j$ with intermediate heavy neutrino\,\cite{Mitra:2016kov} have been performed, which provides an upper limit to the cross section in this channel\,\cite{ATLAS:2023cjo}.
The constraint on heavy neutrino is, on the other hand, less constrained with a lower bound around 200~GeV. Despite the absence of the discoveries, the future runs of the LHC and other future colliders could be key platforms for discovering these particles. We, therefore, consider scenarios containing within the limits of these searches.

The article is organized as follows. A discussion on the heavy-charged gauge boson and its couplings, including the Lagrangian considered, is presented in Section~\ref{sec:lag}. The existing experimental constraints and the choice of benchmark points for further analysis are discussed in Section~\ref{BP_Selection}. The usage of angular variables of the decays of heavy neutrinos is explained in Section~\ref{sec:heavyN}. The discussion on the suggested signal, including the discussion of the employed jet substructure methods and polarization-sensitive variables, is presented in Section~\ref{sec:signal}. The necessary SM backgrounds and discovery potential of the signal are discussed in Sections~\ref{sec:bkg} and \ref{sec:signi}, respectively. The methodology and performance of distinguishing left vs. right-chiral hypotheses are provided in Section~\ref{PolClassify}. Finally, a summary is presented in Section~\ref{sec:summary}.

\section{Heavy Charged Gauge Boson ($W'$)}\label{sec:model}
The coupling of heavy $W'$ bosons to leptons and heavy neutrinos is particularly interesting because it could provide insights into the structure of the extended symmetries and the mechanisms underlying neutrino mass generation.
In the context of the LRSM or similar models with additional gauge groups, the $W'$ boson couples to the right-handed charged leptons and heavy right-handed neutrinos (often denoted as $N_R$). 
Additionally, the models with extra dimensions may also include non-standard $W'$ bosons from Kaluza-Klein excitations, which could exhibit distinct couplings compared to their SM counterparts. For example, in the RS model, the $W'$ bosons can have couplings to heavy left-handed neutrinos and leptons.
Thus, there exist a variety of possible couplings of $W'$ boson with left- or right-handed heavy neutrinos and leptons.

These different types of interactions of $W'$, {\it viz.} interaction to left and right chiral fermions, could lead to distinct experimental signatures that might be observable in high-energy collider experiments or rare decay processes, providing valuable understanding about the nature of new physics beyond the Standard Model.
In this, we focus on the proton-proton collisions at high-energy colliders at the High Luminosity Large Hadron Collider (HL-LHC), where a $W'$ boson could be produced and subsequently decay into a charged lepton and a heavy neutrino. The heavy neutrino could then decay into a charged lepton and a $W$ boson. In the scenario where $W'$ is much heavier than the heavy neutrino, the heavy neutrino produced is significantly boosted, resulting in a very small angle between the decay products of $N$. 
The hadronic decay of $W$ boson thus leads to the final state containing a charged lepton and a fat jet, forming out the decays of boosted heavy neutrino\,\cite{Mattelaer:2016ynf} and containing a lepton inside it.

More importantly, the angular distribution due to the spin alignment of the decay products of heavy $W'$ provides information on the specific coupling of the $W'$ boson to a pair of fermions.
The Lorentz structure at the production and decay vertices sets the distribution of the polarization-sensitive variables at the final state.
For instance, if the $W'$ boson predominantly couples to the left(right)-chiral fermions, depending on the nature of coupling, the resulting decay products will exhibit corresponding left(right)-handed polarization, leading to distinctive asymmetries in their angular distribution\,\cite{Han:2012vk}. These asymmetries can be extracted from the heavy neutrino-induced fat jet, thanks to the advancement of jet substructure techniques and polarization-sensitive jet substructure observables.
Therefore, it is possible to measure the polarization effect experimentally, which might deepen our insights into the nature of the $W'$ boson and the underlying interactions in Beyond Standard Model (BSM) physics. Although the specific nature and strength of these couplings depend on the details of the BSM model, including the gauge symmetries, the presence of mixing between left- and right-handed states, and the mechanisms of symmetry breaking, the method of examining them can be cast in a generic model-independent way by introducing left- and right-handed coupling separately. This is the discussion of the following subsection.

\subsection{A generic Lagrangian for $W'$ to left- and right-handed fermion}
\label{sec:lag}
In this paper, we are not restricting ourselves to a specific BSM scenario or model. Rather, we attempt to characterize the differences between the left and right-handed couplings of heavy vector bosons with the fermions and the possibility of detecting these types of couplings at the HL-LHC. Therefore, we add some additional terms in our SM lagrangian in a model-independent way, considering them as effective couplings. We also consider two additional particles, {\it viz.} a heavy vector boson $W'$, and a heavy neutrino $N$. The effective Lagrangian is, therefore, written as,
\begin{eqnarray}
\label{Lagrangian}
	\mathscr{L}_{\rm new} &=& g'_{\ell L} {W'}_\mu \bar{N}_L \gamma^{\mu} \ell_L + g'_{\ell R} W'_\mu \bar{N}_R \gamma^{\mu} \ell_R + g_{\ell L} W \bar{N}_L \gamma^{\mu} \ell_L + g_{\ell R} W \bar{N}_R \gamma^{\mu} \ell_R \nonumber \\
	&& +\  g'_{q} {W'} (\bar{d}_R \gamma^{\mu} u_R + \bar{d}_L \gamma^{\mu} u_L) + \mathrm{h.c.}
\end{eqnarray}
Here, $g'_{\ell L}$ denotes the coupling strength of the $W'$ to left-handed heavy neutrino and left-handed SM leptons, and $g'_{\ell R}$ is the coupling strength of the same with right-handed heavy neutrino and right-handed SM leptons. Here, $N_L = P_L N$ and $N_R = P_R N$, whereas $P_L$ and $P_R$ are, respectively, the left and right projection matrices. There are two additional left and right couplings called $g_{\ell L}$ and $g_{\ell R}$, which are the couplings of SM $W$ boson with heavy neutrino and SM leptons.
One can easily identify that the left and right-handed couplings of additional vector bosons will portray different polarization information through the final state decay products.

{  In the Lagrangian presented in Eq.~(\ref{Lagrangian}), we have considered simplified scenario describing the effective couplings $g_{\ell L}$, $g_{\ell R}$, $g_{\ell L}'$, and $g_{\ell R}'$. This effective Lagrangian can arise from realistic ultraviolet (UV)-complete models such as LRSM\,\cite{Mohapatra:1974gc,Mohapatra:1974hk,Senjanovic:1975rk,Senjanovic:1978ev}, which provides a natural framework for explaining the origin of parity violation in the SM as a consequence of spontaneous symmetry breaking. Additionally, the Universal Seesaw Model with the extension of the LRSM\,\cite{Dey:2022tbp}, which explains the mass generation of all fermions (quarks and leptons) using a seesaw mechanism, is another realistic model.
In these LRSM scenarios, the SM electroweak is extended to $SU(2)_L \times SU(2)_R \times U(1)_{B-L}$, naturally introducing right-handed weak interactions 
alongside the usual left-handed ones
with additional right handed heavy gauge bosons, $W'$ and $Z'$. As a direct consequence of promoting right-handed fermions to doublets, right handed neutrinos are introduced, which couple naturally to the $W'$.
After mass mixing in both the gauge and fermionic sectors, the SM like $W$ and new $W'$ can couple with those heavy neutrinos $N$ and charged leptons.

On the other hand, couplings between $W'$ and left-handed neutrinos can naturally arise in the models with extra spatial dimentions\,\cite{Arkani-Hamed:1998jmv,Randall:1999ee}. For example, the scenario proposed in Ref.\,\cite{Grossman:1999ra} for Dirac neutrino mass generation is formulated within the framework of the RS model with a non-factorizable geometry. In brief, the RS model with orbifolding features two branes: one associated with the Planck scale (UV-brane) and the other corresponding to the electroweak scale (the TeV- or visible-brane). The SM fields are localized on the TeV brane, where their mass scales are suppressed by a warp factor with respect to the UV-brane, naturally placing them in the TeV mass scales. Any particle, which propagates through the full $4+1$ dimensional bulk, manifests on the brane through its Kaluza-Kein excitaions. In the setup in Ref.\,\cite{Grossman:1999ra}, massless neutrinos are allowed to access the bulk, leading to KK excitations whose mixing with the zero modes generates small neutrino masses. If the electroweak gauge boson $W$ is also allowed to access the bulk, it brings in its own KK excitations ($W'$). Thus the coupling between the heavy left-handed neutrinos and the heavy $W'$, as presented in Lagrangian~[Eq.~(\ref{Lagrangian})], arises naturally.

In the following, we discuss the production and decays of $W'$ considering the simplified set up as given in Lagrangian~[Eq.~(\ref{Lagrangian})].} The heavy vector boson can be generated at the LHC through $pp$-collision via the coupling $g'_{q}$. After production, it can decay to a heavy neutrino and a charged lepton, followed by the heavy neutrino decay to SM $W$-boson and leptons. When we keep $g_{\ell L}$ and $g'_{\ell L}$ non-zero with right-handed couplings at zero, it will show the distribution of the left-handed polarized state of the heavy neutrino coupling to $W'$. On the other hand, when $g_{\ell R}$ and $g'_{\ell R}$ are non-zero with zero left-handed couplings, it will show the distributions of the right-handed polarized state of the heavy neutrino coupling to $W'$. These parameters, therefore, set the angular distributions corresponding to the left and right-handed hypotheses. We now move on to the current experimental constraints on these parameters.

\vspace{-4mm}
\subsection{Experimental Constraints and Benchmark Points Selection:}
\label{BP_Selection}
There are many existing experimental searches that could potentially constrain the parameters of the given Lagrangian in Eq.~(\ref{Lagrangian}). These experimental constraints mainly arise from searches for heavy non-standard states in existing collider experiments. The non-observation of any direct signal at the LHC has put stringent constraints on the mass of the heavy right-handed gauge boson, $W'$, and the heavy neutrinos, $N$.

The production cross section of heavy resonances had been strongly constrained by the search associated with them decaying to the di-lepton/di-jet final state at the LHC. The searches for $W'$ in the $\ell^{\pm} N$ channel, where $\ell$ is either $e$ or $\mu$ and $N$ is the heavy neutrino, impose the primary experimental constraint on our case. Depending on the mass difference of $W'$ and $N$, the final state is either one charged lepton plus a fat jet containing a high $p_T$ lepton\,\cite{CMS:2021dzb} or two leptons and two jets\,\cite{ATLAS:2019fgd}. For the models with the same left and right-handed gauge couplings ($g_R$=$g_L$), such as the Sequential Standard Model\,\cite{Altarelli:1989ff} and Left-Right symmetric model, the latest CMS search\,\cite{CMS:2021dzb} places a lower limit on $W'$ that excludes its masses below 4.7 TeV and 5.0 TeV for the electron ($e^{\pm}$) and muon ($\mu^{\pm}$) channels, respectively.

However, for some particular models, the ATLAS\,\cite{ATLAS:2019fgd} di-jet search places a lower bound of 4 TeV on $W'$ mass with SM-like couplings, considering 50\% branching ratio in di-jet mode. For various models, these limits vary according to the branching ratio of $W'$ in the di-jet and the $\ell^{\pm} N$ channels. In our case, the $W'$ boson can only have two decay modes, as we have written down an effective coupling in addition to the Standard model. In particular, $W' \rightarrow jj$ with $\sim 75\%$ branching ratio and $W' \rightarrow N \ell^{\pm}$ with $\sim 25\%$ branching ratio. Considering the fact that the mass of $W'$ ($M_{W'}$) is significantly more than the mass of the heavy neutrino ($M_N$), the heavy neutrino $N$ has only one decay channel through the process $N \rightarrow W^\mp \ell^{\pm}$ with $100\%$ branching, with $~30\%$ branching in each of $W^\mp e^{\pm}$ and $W^\mp \mu^{\pm}$ channels. Since we are taking a more model-independent approach without introducing a new gauge group, the masses of our heavy boson states are independent of any additional gauge couplings or VEVs in our models. 
As a result, the couplings found in Eq.~(\ref{Lagrangian}) are independent of the model and unaffected by gauge or Yukawa couplings. Taking that into account, we have left out any theoretical restrictions here. Furthermore, this also allows us to take the mass of the $W'$ boson lower than 4~TeV by adjusting the coupling parameters in such a way that the production cross section in a particular channel remains below the observed upper limit.

{\bf Choice of parameter space:}
We select a few benchmark points for our analysis, where the mass of the heavy neutrino is chosen to lie between $200$ and $400$~GeV, while the mass of the $W'$ is varied in the range of $3$ to $6$ TeV. These mass ranges are also popular choices among experimental collaborations such as CMS\,\cite{CMS:2021dzb} and ATLAS\,\cite{ATLAS:2019fgd} at the 13 TeV LHC. The coupling values for our benchmark points are listed in Table~\ref{Tab:benchmark}, as specified in Eq.~(\ref{Lagrangian}). These couplings have been verified to ensure that the total production cross section remains below the 95\% upper limit set by CMS\,\cite{CMS:2021dzb} and ATLAS\,\cite{ATLAS:2019fgd}. {  Additionally, the benchmark points have also been checked to satisfy the CMS heavy $W'$ search in the $2\ell+2j$ channel\,\cite{CMS:2021dzb} at the 13~TeV LHC with an integrated luminosity of 138~fb$^{-1}$. We used the publicly available and validated recasting\,\cite{Frank:2023epx,DVN/UMGIDL_2023}, implemented within {\tt MadAnalysis5} framework\,\cite{Conte:2012fm,Conte:2014zja, Conte:2018vmg}, to reinterpret the CMS search results. The details of this result are provided in Appendix~\ref{AppendixA}.}

\begin{table}[h]
\begin{center}
	\begin{tabular}{|c||c|c|c|c|c|}
		\hline
		Benchmarks Points & $M_{W'}$ (TeV) & $M_N$ (GeV) & \qquad $g'_q$\qquad\qquad & \quad$g'_{\ell L}/g'_{\ell R}$\quad\quad & \quad $g_{\ell L}/g_{\ell R}$ \quad\quad
		\\
		\hline
		BP1 & 3.0 & 200.0 & 0.03 & 0.08 & 0.1  
		\\
		BP2 & 3.0 & 400.0 & 0.03 & 0.08 & 0.1  
		\\
		BP3 & 4.5 & 200.0 & 0.15 & 0.45 & 0.1  
		\\
		BP4 & 4.5 & 400.0 & 0.15 & 0.42 & 0.1  
		\\
		BP5 & 6.0 & 200.0 & 1.4 & 1.4 & 0.1  
		\\
		BP6 & 6.0 & 400.0 & 1.4 & 1.4 & 0.1  
		\\
		\hline
	\end{tabular}
\end{center}
\vspace{-5mm}
\caption{Details of benchmark points.}
\label{Tab:benchmark}
\end{table}

As outlined in the Lagrangian [Eq.~(\ref{Lagrangian})], the left- and right-handed couplings are governed by the parameters $g_{\ell L}$, $g_{\ell R}$, $g'_{\ell L}$, and $g'_{\ell R}$. For each benchmark point, the specific values of these couplings are provided in the last two columns of Table~\ref{Tab:benchmark}. Each benchmark point includes two scenarios: one with only left-handed couplings nonzero and the other with only right-handed couplings nonzero. Our convention is as follows: when the left-handed couplings are nonzero ($g_{\ell L}, g'_{\ell L} \neq 0$, and $g_{\ell R}, g'_{\ell R} = 0$), the scenario is labelled as `LH'. Conversely, when the right-handed couplings are nonzero ($g_{\ell R}, g'_{\ell R} \neq 0$, and $g_{\ell L}, g'_{\ell L} = 0$), the scenario is labelled as `RH'. Benchmark points are denoted accordingly, with the appropriate `LH' or `RH' abbreviation appended. For instance, BP1-LH represents benchmark point BP1 with $g_{\ell R} = g'_{\ell R} = 0$.

\subsection{Role of Heavy Neutrinos in Polarization Measurements:}\label{sec:heavyN} 

When the heavy neutrino is highly boosted, the chirality becomes almost the same as its helicity (or polarization). The chiral nature of $W'$ coupling to a charged lepton and a heavy neutrino ($N$) can be studied through the decay products of $N$ since the polarization of $N$ depends on the nature of its coupling to $W'$. For the chosen benchmark points considering various experimental constraints, the heavy neutrino $N$ decays to $\ell W$ with 100\% branching fraction. In the hadronic decay mode of $W$ boson, it then decays to an up-type ($u$) and a down-type ($d$) quark anti-quark pair. This decay mode is very similar to the decay mode of the top quark in the SM, wherein the top quark decays to a bottom quark and a $W$ boson with almost 100\% branching ratio. Therefore, the polarization study of the heavy neutrino can be performed in a similar manner as the top quark polarization study\,\cite{Godbole:2019erb}. In such a decay mode, the angular distribution of the three decay products ($f = \ell, u, d$) of $N$ with polarization $\mathcal{P} \in (-1,1)$ can be written as\,\cite{Jezabek:1988ja}
\begin{equation}
\frac{1}{\Gamma} \frac{d\Gamma}{d\cos\theta_f} = \frac{1}{2}(1 +  A_f \mathcal{P} \cos\theta_f ),
\end{equation}
where $\theta_f$ is the angle between decay product $f$ and spin direction of $N$ in its rest frame. The coefficient $A_f$ is called spin analyzing power corresponding to the decay product $f$. The spin analyzing power provides the measure of forward-backward asymmetry in the rest frame of the decaying particle. Suitably constructed variables from decay products with high analyzing power could provide higher distinguishing features among various polarized states. 

Assuming that the decayed fermions are massless, the spin analyzing power depends on two things: (a) the nature of the chiral coupling, {\it viz.} $V+A$ (RH) and $V-A$ (LH), and (b) the mass ratio $M_W/M_N$\,\cite{Jezabek:1994zv}. 
The expressions for various $A_f$ and their values corresponding to the two chosen mass points of heavy neutrino $N$ have been listed in Table~\ref{Tab:spin-analyzing}. Actually, the analyzing power of $u$ quark for LH is simply negative of the analyzing power of $d$ quark for RH cases. Although the down (up)-type quark 
offers the highest analyzing power in the LH (RH) case, its utility is limited. This is because the quarks are never detected freely, and they are detected as jets. Isolating down- or up-type quark-induced jets is not efficiently achievable in the busy environment at the LHC. The task is further compounded in our signal topology, where these quarks come together to form a fat jet. However, isolating the lepton, which also offers reasonably good analyzing power between $-0.5$ to $-0.85$, as a subjet inside the fat jet is a little easier. We will discuss the polarization-sensitive variables with this subjet and other subjets in the next section, along with the discussion of the signal topology. 

\begin{table}[t]
\renewcommand{\arraystretch}{1.6}
\begin{center}
	\begin{tabular}{|c|c|c|c|}
		\hline 
		& $x = \dfrac{M_W^2}{M_N^2}$ & $M_N = 200$ GeV & $M_N = 400$ GeV  
		\\[7pt]
		\hline 
		$A_\ell^{\rm LH} = A_\ell^{\rm RH}$ & $-\dfrac{1-2x}{1+2x}$ & $-0.51$ & $-0.85$  
		\\
		$A_u^{\rm LH} = - A_d^{\rm RH}$ & $1 - \dfrac{12x(1-x+x\ln x)}{(1-x)^2 (1+2x)}$ & $-0.34$ & 0.60   
		\\
		$A_d^{\rm LH} = - A_u^{\rm RH}$ & 1  & 1.0 &  1.0 
		\\
		\hline
	\end{tabular}
\end{center}
\vspace{-5mm}
\caption{Spin analysing power of different decay products of heavy Neutrino $N$ for two different mass points. The spin analyzing power for $W$ boson is related to that of the lepton as $A_W = - A_\ell$.}
\label{Tab:spin-analyzing}
\end{table}

\section{Analyses and Results:}
\label{sec:analysis}
In this section, we present a comprehensive signal-background analysis for a final state involving a fat jet and an isolated lepton at the HL-LHC at a centre-of-mass energy of $\sqrt{s} = 14$~TeV, incorporating detector simulation. As our focus is on studying this process under HL-LHC conditions, we assume a projected integrated luminosity of $\mathcal{L} = 3$~ab$^{-1}$.

\subsection{Signal}\label{sec:signal}
\begin{figure}
\centering
\includegraphics[width=0.5\textwidth]{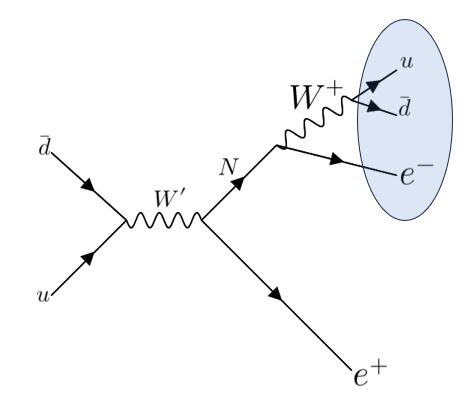}
\caption{Feynman diagram showing the considered signal process: production of a heavy neutrino $N$ via decay of $W^\prime$. $N$ consequently decays through a 3-body process into two quarks and a lepton. The decay products of $N$ are highly collimated, forming a fat jet with a distinct substructure containing both leptonic and hadronic components.}
\label{fig:feyndiag} 
\end{figure}

The signal process we consider involves the production of a heavy neutrino ($N$) via the reaction $p p\rightarrow W' \rightarrow N \ell$ at the LHC. The neutrino then decays through $N\rightarrow\ell q\bar{q'}$, {  primarily through on-shell $W$ boson.}
Figure~\ref{fig:feyndiag} depicts the Feynman diagram corresponding to the signal process under consideration. {  The heavy neutrino $N$ can also decay to the same final state through off-shell $W'$. However, the additional contribution for such a scenario is about a percent for the chosen benchmark points, and hence, we did not consider this scenario in our analysis.}
In our scenario, the $W'$ is much heavier compared to its decay product $N$. Therefore, the resulting heavy neutrino $N$ is produced with a high boost.
This momentum boost causes the three decay products of $N$ to be found in a narrow solid angle\footnote{  A fraction of events, in which all the decay products cannot come within a small solid angle, will exhibit a distinct $2\ell+2j$ final state. As described previously, this scenario has been explored through the CMS and ATLAS collaborations. We focus instead on the complementary phase space.}.
When these decay products are reconstructed in the detector with a large-radius jet algorithm, they become clustered into a single, large-radius jet, also referred to as a fat jet.

The fat jet formed from $N$'s decay possesses a unique three-pronged substructure. Two of the prongs correspond to the hadronic decays from the $W$ boson decay, while the third prong corresponds to the lepton produced in the decay of $N$. Due to the high boost of $N$, this lepton is contained within the radius of the fat jet, rather than being identified as an isolated lepton in the detector. This leptonic component within the fat jet is atypical for standard QCD jets, which are generally purely hadronic. Also, the fat jets arising from the particles in the SM do not usually contain lepton inside them. Therefore, this feature provides a distinguishing feature that helps separate this signal from background processes. Overall, this process yields a clean and identifiable final state that can be effectively differentiated from SM backgrounds.

To simulate this signal process, we employ a multi-step approach that accounts for the entire chain from parton-level event generation to the reconstructed final states as they would appear in the detector. First, we have implemented the model in {\tt FeynRules2.0}\,\cite{Alloul:2013bka} and generated the Universal FeynRules Output (UFO)\,\cite{Degrande:2011ua} files. Then, \texttt{MadGraph5\_aMC@NLO}\,\cite{Alwall:2014hca} is used to generate the hard scattering events at $\sqrt{s} = 14$~TeV using the UFO files.
Next, these events are passed through \texttt{PYTHIA-8.3}\,\cite{Sjostrand:2006za,SJOSTRAND2008852}, which handles the parton showering and hadronization, producing hadronic final states keeping multi-parton interaction, initial- and final-state radiation switched on. Finally, the events are processed with \texttt{Delphes-3.5.0}\,\cite{deFavereau:2013fsa} for detector simulation.
We have taken pile-up effects into consideration, assuming an average of 50 interactions per bunch crossing to make it more realistic. We then used \texttt{PUPPI}\,\cite{Bertolini:2014bba} module implemented with {\tt Delphes} to mitigate the effects of pile-up.
Jet reconstruction was performed in \texttt{Fastjet 3.4.2}\,\cite{Cacciari:2011ma} using the \texttt{Anti-$\mathrm{k_T}$} algorithm with a radius parameter of $R= 0.8$. The {\tt Delphes}-implemented PF Candidates within $|\eta| < 2.5$ and having at least two leptons were used as input to the jet clustering algorithm. Jets having a transverse momentum threshold of $p_T^j>150$ GeV are considered for further analysis. We $p_T$-sort the jets in descending order and select the two highest $p_T$ jets to determine which one corresponds to the fat jet containing a lepton and which one represents the isolated lepton. We employ the following step-by-step methods to identify the isolated lepton, the fat jet and the three subjets, including the lepton subjet, inside the fat jet.
\begin{itemize}
\item We first iterate over the constituents of each of the two leading jets to find the highest $p_T$ lepton in each jet. Once a lepton has been identified within a jet, we calculate the fraction of the jet's $p_T$ carried by that lepton.
\item If the lepton’s $p_T$ exceeds 150 GeV and it carries more than 85\% of the jet’s total $p_T$, we classify this jet as the isolated lepton.
\item The other large radius jet that does not meet the isolated lepton criteria as per the previous point is identified as the fat jet corresponding to the heavy neutrino.
This fat jet is expected to have a unique substructure, featuring both hadronic components from the decay of a $W$ boson and a leptonic component from the decay of the heavy $N$.
\item To further refine the structure of the fat jet, we apply the Soft Drop grooming\,\cite{Larkoski:2014wba} with $z_{\rm cut} =$ 0.1 and $\beta = 1.0$, to remove soft and wide-angle radiation.
\item We then find the 3 subjets inside the soft-dropped fat jet using N-subjettiness\,\cite{Thaler:2010tr}. We analyze the constituents of the subjet to locate the specific subjet containing the embedded lepton.
\end{itemize}

We have generated the signal process for six distinct benchmark points, as described in Section~\ref{BP_Selection}, for both right- and left-handed heavy neutrinos $N$. In our process, the neutrino $N$ decays into a three-pronged fat jet ($N \rightarrow \ell j j $). Drawing an analogy with the top-quark decay ($t \rightarrow bjj$), as discussed in Section~\ref{sec:heavyN}, we have modified the definitions of the variables $z_k$ and $z_b$\,\cite{Krohn:2009wm, Godbole:2019erb}, which are suggested for the top quark polarization studies.

The tagged jet, associated with the heavy neutrino $N$, contains three subjets, with one of these subjets being a lepton $\ell$ embedded inside it.
Among the possible combinations of these subjets, we identify the pair with the smallest $k_T$ distance. The $k_T$ distance $d_{ij}$ between subjets $i$ and $j$ is defined by
\begin{equation}
d_{ij} = \min(p_{Ti}^2, p_{Tj}^2) R_{ij}^2,
\end{equation}
where $R_{ij}^2 = (y_i-y_j)^2 + (\phi_i-\phi_j)^2$. The energy fraction of the harder subjet $j_k$ from this pair, which has the smallest $k_T$ distance, is denoted as $z_k$:
\begin{eqnarray}
z_{k} & = & \frac{\max\left(E_{i},E_{j}\right)}{E_{N}}
\text{, for which} \: d_{ij} \: \text{is minimum}
\label{eq:zk}
\end{eqnarray}
and $E_N$ is the energy of the jet, coming from $N$.
The second variable we use is the energy fraction of the lepton-like subjet, $z_\ell$ can be defined as,
\begin{equation}
z_\ell = \frac{E_\ell}{E_N}.
\label{eq:zl}
\end{equation}
Additionally, we have constructed one more polarization-sensitive variable $z_\theta$. Inspired by the technique in\,\cite{De:2020iwq,Dey:2021sug}, which has been used to measure $W$ boson polarization by taking the ratio of the energy difference between the two subjets of $W$ fat jet to the $W$ fat jet three momentum. This variable constructs a proxy for the decay polar angle in the rest frame of $W$\,\cite{De:2020iwq}. We apply a similar approach by using the energy difference between two of the three subjets from the $N$ decay, normalized by the momentum of $N$.
Out of the two subjets one is always the lepton subjet. The other is chosen to be the one which produces the lowest invariant mass with the lepton subjet. Such a variable is also suggested for the top quark polarization measurement with respect to the $b$ the subjet inside the top fat jet\,\cite{Godbole:2019erb}.
Mathematically, the variable $z_\theta$ is expressed as:
\begin{equation}
z_\theta = \frac{ |E_{\rm subj}^\ell - E_{\rm subj}^h| }{|\vec p_{N}|},
\label{eq:ptheta}
\end{equation}
where $E_{\rm subj}^{\ell/h}$ represent the energies of the selected leptonic and hadronic subjets (from the pair with the smallest invariant mass), and $\vec p_{N}$ is the three momentum of the $N$ fat jet. The variables $z_k$, $z_\ell$ and $z_\theta$ can be defined in a similar way at the parton-level generated by {\tt MadGraph5}, where the subjets should be replaced by lepton or quarks appropriately.

\begin{figure}[h]
\centering
\includegraphics[width=\linewidth]{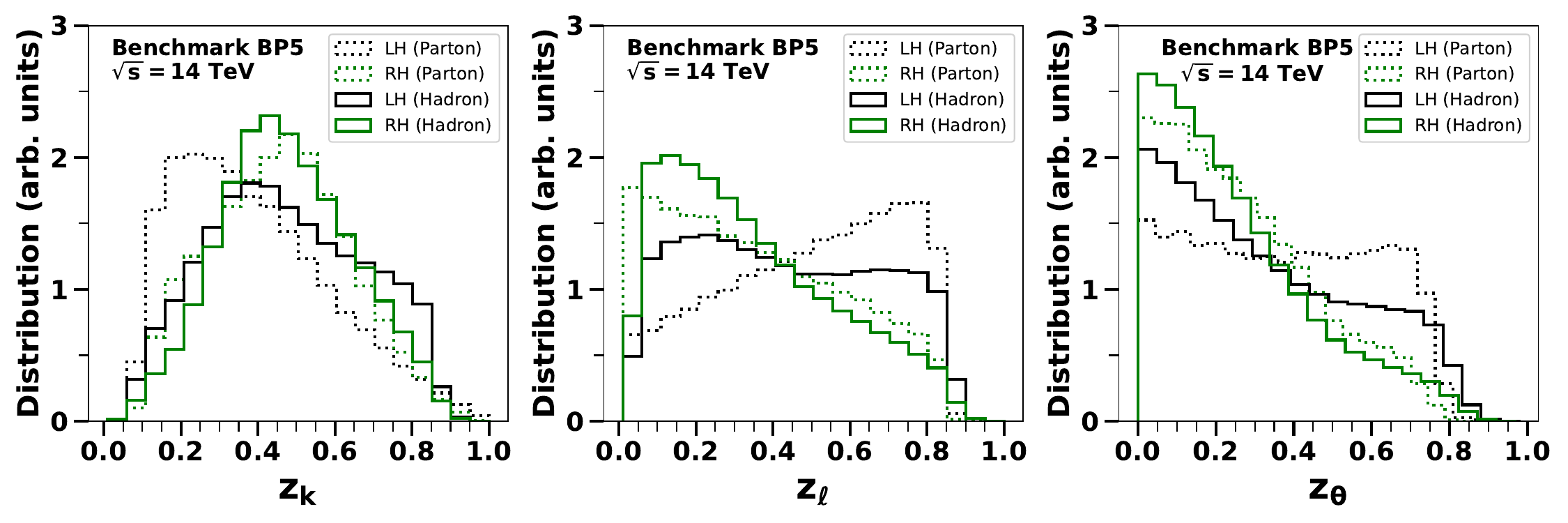}
\includegraphics[width=\linewidth]{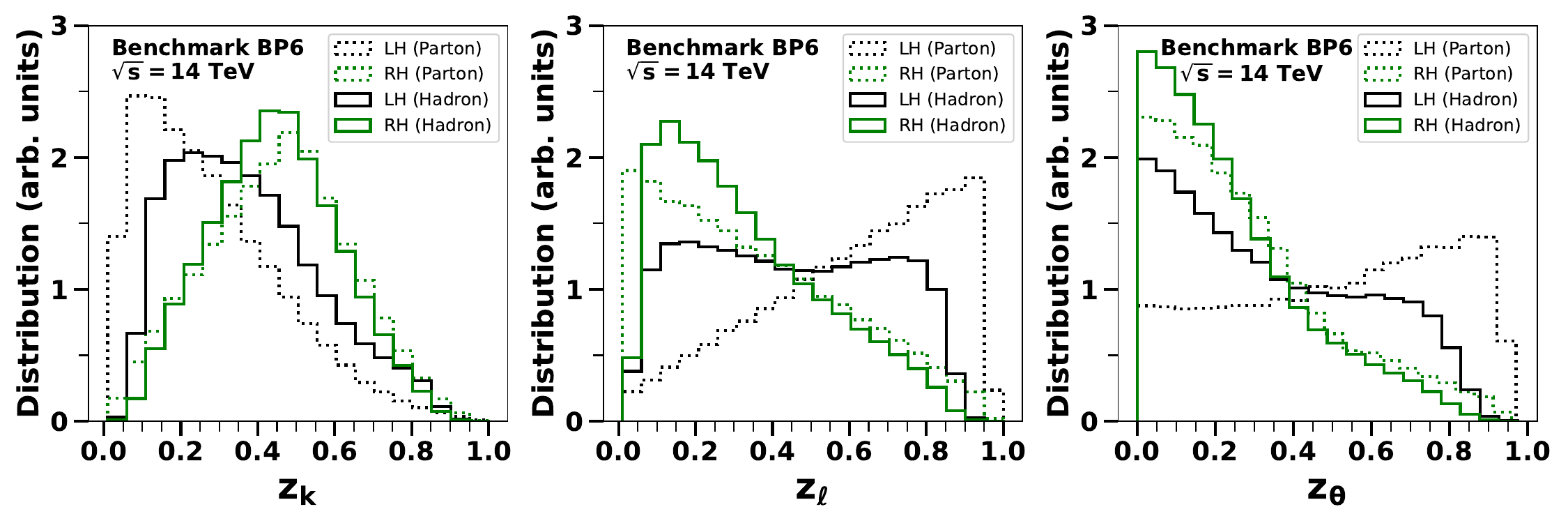}
\caption{Comparison of the distributions of $z_k$, $z_\ell$, and $z_\theta$ for left-handed (LH) and right-handed (RH) helicity configurations of massive neutrino $N$ in both parton-level and hadron-level analyses of the signal process for Benchmarks BP5 (top row) and BP6 (bottom row). Parton-level distributions are represented by dashed lines, while hadron-level distributions are shown with solid lines.}
\label{fig:partonDist_BP5}
\end{figure}

In Fig.~\ref{fig:partonDist_BP5}, we present the distributions of the $z_k$, $z_\ell$, and $z_\theta$ variables for the benchmark points BP5 (top) and BP6 (bottom),
respectively. The distributions are shown for LH and RH configurations by black and green curves, respectively. In each plot, the solid lines labeled as `Hadron' represent the hadron-level distribution after detector simulation and the corresponding dashed lines labeled as `Parton' represent the parton-level distribution. As expected, clearer separations between LH and RH are seen in parton-level distributions, and the hadron-level distributions are smeared off due to the hadronization and detector effects. The distributions for the other four benchmark points are moved to Appendix~\ref{AppendixC} to keep the section tidy.

The $z_k$ distribution representing the energy fraction of the hardest subjet of the least $k_T$ distant pair suggests that LH configurations have relatively smaller energy fractions, while RH configurations are more peaked near 0.5, implying more compact energy sharing among subjets. Similarly, the $z_\ell$ distribution shows that the energy fraction carried by the lepton tends to be higher in the LH case, whereas RH events exhibit a sharper drop-off, indicating that energy is more concentrated in the quark subjets. In line with the concept outlined in\,\cite{De:2020iwq}, the $z_\theta$ variable serves as a rough indicator of the angular correlation between the leptonic and hadronic subjets. The $z_\theta$ distribution for the left-handed (LH) configuration exhibits a broader spread, indicative of a more relaxed angular separation. Conversely, the right-handed (RH) configuration displays a narrower distribution, suggesting tighter angular correlations between the subjets.

Based on these observations, we can say that $z_\ell$ and $z_\theta$ are more robust variables for distinguishing between the helicity states of the neutrino $N$. This is because these two observables involve the lepton, which has a relatively good analyzing power. Additionally, these two variables retain good discriminating power even at the hadron-level because the lepton subjet is relatively simpler to isolate compared to the hadron subjet in the busy environment inside the fat jet. While $z_k$ is less effective than  $z_\ell$ and $z_\theta$, it still provides valuable complementary information. Thus, we incorporate $z_k$ alongside $z_\ell$ and $z_\theta$ in our analysis. By utilizing a multivariate analysis such as Boosted Decision Tree (BDT)\,\cite{BDT_Ref1}, we can effectively capture the distinct relationships among these variables and improve the overall separation power between the left- and right-handed polarization states of $N$.
We will take up this in Section~\ref{PolClassify} to utilize $z_\ell$, $z_\theta$ and $z_k$ within the BDT framework to achieve a comprehensive and robust classification of the polarization states.

\vspace{-2mm}
\subsection{Backgrounds}\label{sec:bkg}\vspace{-2mm}
In the search for our aforementioned signal containing one fat jet and one high $p_T$ isolated lepton, several SM backgrounds need to be considered. The QCD multijet background, due to its high production cross section, is prevalent in any hadron collider. We have taken events with a 2-jet final state at the leading order (LO). Top pair production ($t\bar{t}$) is another dominant background, particularly in its semi-leptonic decay mode, which can produce a fat jet from the hadronic decay of one top quark and a high $p_T$ lepton from the leptonic decay of the other top. Additionally, vector boson production in association with jets ($V+$jets, where $V = W/Z$) can generate final states with leptons and jets. Furthermore, diboson production ($VV$) can also contribute, as one of the bosons can decay hadronically, forming a fat jet, while the other decays leptonically, creating a final state that closely resembles our signal. 

We have followed a similar approach for simulating the background events as for the signal events.
For the $V+$jets background, we considered the production of vector bosons ($V = W/Z$) in their leptonic decay modes and with one or two associated jets {  ($V+1,2j$) at the MG5 parton-level at the LO, which were then showered using {\tt Pythia8}. We then used MLM merging~\cite{MLM,Mangano:2006rw,Mrenna:2003if,Hoeche:2005vzu,Alwall:2007fs} prescription with $q_{\rm cut} = 30$~GeV to take care of the overcounting during to combine exclusive sample with different jet multiplicities. Moreover, in this sample, off-shell photon contributions are possible; however, we have checked that their impact in the phase space of interest is negligible.
For the diboson background, we simulated events with up to one additional jets ($VV+0,1j$) at the LO. The MLM merging scheme with $q_{\rm cut} = 30$~GeV has been applied in this case also. All the backgrounds have been generated at the LO. We then multiplied the cross sections by appropriate $k$ factors to make up for the corrections at the next-to-next-to-leading-order (NNLO) for $t\bar t$ samples, and at the next-to-leading-order (NLO) for $V+$jets and $VV+$jets. The corresponding $k$-factors are 1.6~\cite{Czakon:2013goa}, 1.36~\cite{Kim:2024ppt}, 1.41~\cite{Kim:2024ppt} for $t\bar t$, $V+$jets, and $VV+$jets, respectively. }To populate the signal region efficiently, we require the parton-level centre-of-mass energy $\sqrt{\hat{s}} >$ 1.5 TeV for $t\bar{t}$, $V+$jets and $VV+$jets backgrounds and $\sqrt{\hat{s}} >$ 1.7 TeV for QCD backgrounds. Additionally, we also require the transverse momentum of the two partons $p_T^{j_1,j_2} >$ 100 GeV and the scalar sum of the transverse momentum of the partons $H_T >$ 1.7 TeV for QCD background and lepton $p_T^\ell >$ 100 GeV for $V+$jets background.
This cuts at the parton level would not underestimate the background since our signal requires the production of $W'$, which has been considered at least 3~TeV for all the BPs. We then performed detector simulation with \texttt{Delphes} and the jet clustering and analysis of jet substructure using \texttt{FastJet}. In the subsequent paragraphs, we discuss how these backgrounds can be suppressed through kinematic cuts to enhance the signal sensitivity.

\vspace{-7mm}
\subsection{Significances}
\begin{figure}[h]
\centering
\includegraphics[width=\linewidth]{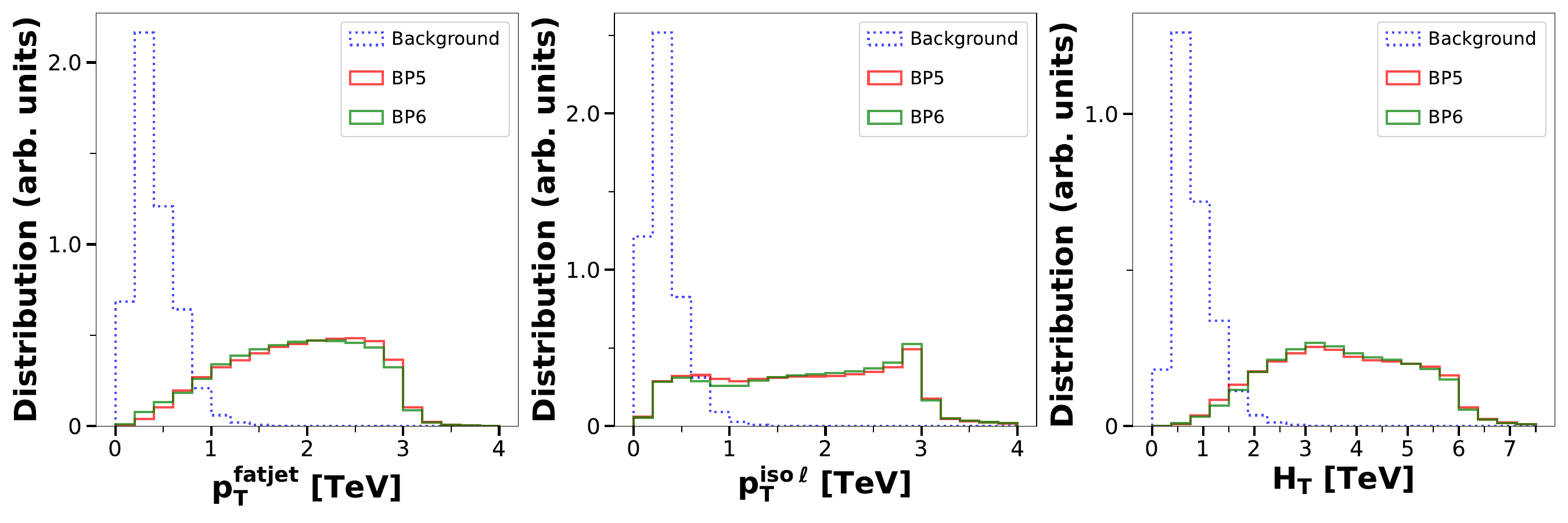}
\caption{Distributions of key kinematic variables for the signal and background processes in Benchmark BP5 and BP6. \textit{Left:} $p_T$ distribution of the fat jet, \textit{Center:} $p_T$ distribution of the isolated lepton, \textit{Right:} distribution of the scalar sum of jet transverse momenta $H_T$}\vspace{-2mm}
\label{fig:KinVar_BP56_Set1}
\end{figure}

\vspace{-3mm}\label{sec:signi}
We plot the distribution of $p_T$ of fat jet and the isolated lepton, and $H_T$ (scalar sum of the $p_T$'s of all reconstructed jets and isolated lepton) in Fig.~\ref{fig:KinVar_BP56_Set1}. From the figure, we see the signal distribution extends into the higher $p_T$ and $H_T$  region compared to the background, indicating that the signal events are associated with a higher-energy environment. This motivates us to require $H_T >$ 2000 GeV, ensuring that we select events with a high-energy environment, which is characteristic of heavy $W'$ production. We also impose a partonic centre-of-mass energy cut, $\sqrt{\hat{s}} > 2000$ GeV, to filter out lower-energy background processes. At the parton level, $\sqrt{\hat{s}}$ represents the total energy in the center-of-mass frame of the colliding partons, but at the hadron level, we reconstruct it by using the total visible hadrons in the event. Additionally, looking at the $p_T$ distributions in Fig.~\ref{fig:KinVar_BP56_Set1}, we apply a stringent cut on the fat jet transverse momentum, $p_T^{\text{fatjet}} > 750$ GeV, which helps to isolate the high-energy jets resulting from the heavy neutrino decay, and a similar cut on the isolated lepton transverse momentum, $p_T^{\text{iso}\:\ell} > 500$ GeV, ensuring that only events with energetic leptons are considered. These cuts help us significantly reduce the number of backgrounds.

To further enhance the signal significance over the aforementioned backgrounds, we focus on two key jet substructure-based variables that exhibit distinct distributions for the signal compared to the SM backgrounds. The variables `$LSF$' and `$LMD$', also used in CMS searches to suppress QCD backgrounds, are particularly effective in the analyses of fat jets containing lepton embedded inside them.
To calculate the $LSF$ and $LMD$ variables, all final-state particles in the event, including leptons, are clustered into a fat jet, and then three sub-jets are found with $N$-subjettiness method, as described in Section~\ref{sec:signal}.
We then identify the sub-jets associated with high $p_T$ leptons, and $LSF$ for the lepton is calculated as the ratio of the $p_T$ of the lepton, ${p_T}_\ell$ to the $p_T$ of the associated subjet, ${p_T}^{s_j}_\ell$:
\begin{equation}
LSF = \frac{{p_T}^\ell}{{p_T}_{\rm subj}^\ell}
\end{equation}
The $LMD$ variable is constructed in a similar manner to the $LSF$ and is defined as:
\begin{equation}
LMD = \frac{m^2_{\text{subj}-\ell}}{m^2_{\rm subj}}
\end{equation}
where $m^2_{\rm subj}$ represents the square of the invariant mass of the subjet associated with the lepton, and $m^2_{\text{subj}-\ell}$ denotes the square of the invariant mass of the same subjet with the lepton subtracted.
\begin{figure}[h]
\centering
\includegraphics[width=0.8\linewidth]{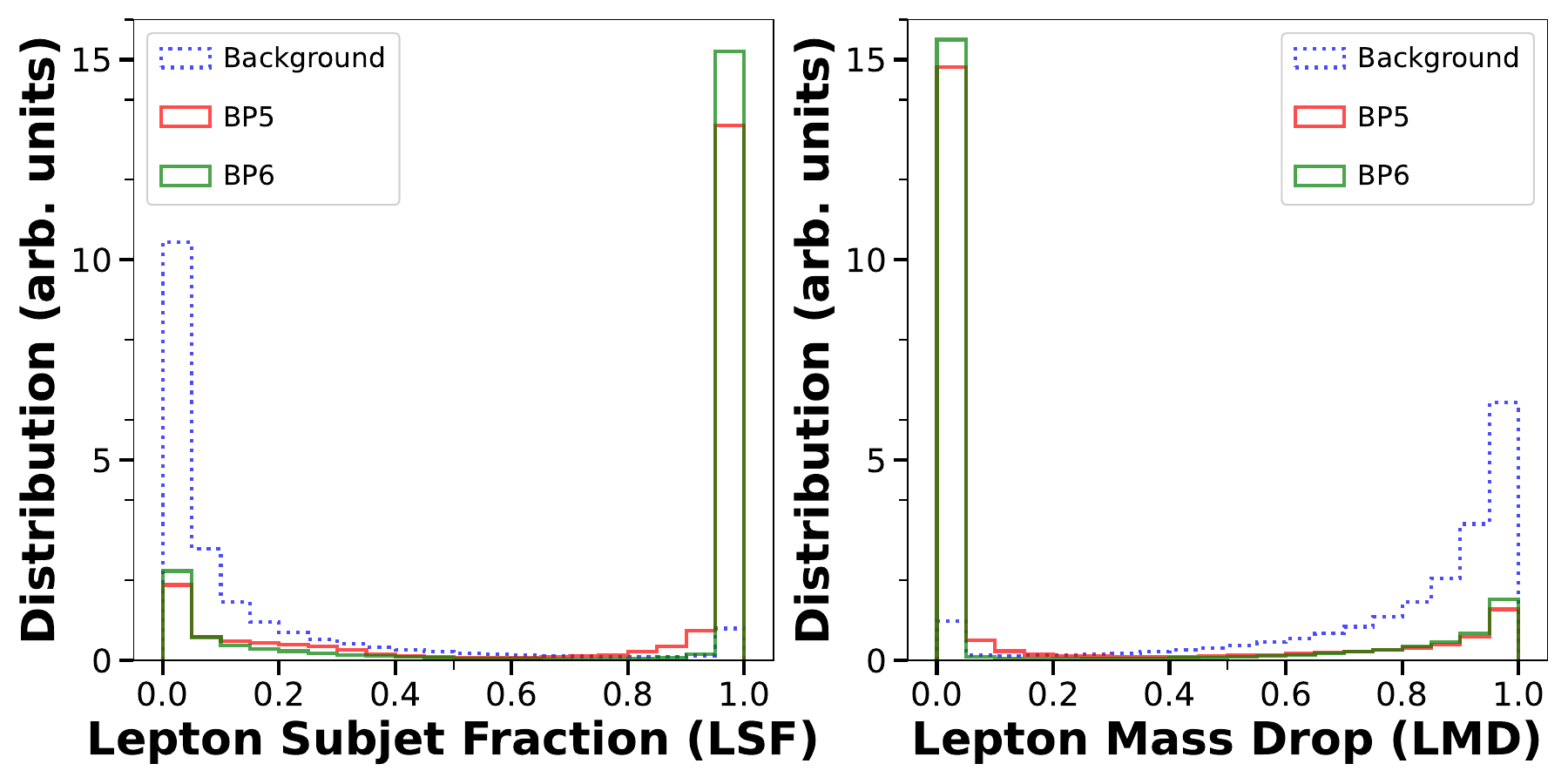}
\caption{Distributions of jet substructure variables for the signal and background processes in Benchmark BP5 and BP6. \textit{Left:} distribution of $LSF$, which measures the fraction of transverse momentum carried by the lepton within the fat jet, and \textit{Right:} distribution of $LMD$, which indicates the mass drop within the lepton-subjet.} \vspace{-3mm}
\label{fig:KinVar_BP56_Set2}
\end{figure}
We have presented the distribution for the $LSF$ and $LMD$ variables in Fig.~\ref{fig:KinVar_BP56_Set2}, showcasing their distinct distributions for the signal compared to the SM backgrounds. Based on the cut values utilized in Ref.\,\cite{Dey:2022tbp} and guided by the observed distributions, we impose the following requirements: $LSF > 0.9$, and $LMD < 0.05$. The plots in Fig.~\ref{fig:KinVar_BP56_Set1} and~\ref{fig:KinVar_BP56_Set2} correspond to two specific benchmark points BP5 and BP6. For completeness, plots for other benchmark points are also provided in the Appendix~\ref{AppendixC}. One can easily guess that these two jet substructure-based variables, along with the kinematic variables, would significantly enhance signal significance. 

We are now in a position to discuss the discovery potential of the signals at the 14 TeV HL-LHC. Our approach is cut-based analysis, in which we have evaluated the signal significance by systematically analyzing the impact of each applied cut on both signal and background events.
By assessing the performance of cuts on key variables, we have identified the most effective strategies for isolating signal regions where high signal significances are achieved.

\begin{table}[ht]
\centering
\begin{tabular}{|>{\arraybackslash}p{2.5cm}|>{\centering\arraybackslash}p{2.7cm}|c|c|c|}
	\hline
	\multirow{2}{*}{\quad\textbf{Cutflow}} & \multirow{2}{*}{\textbf{Initial}} & $\mathbf{H_T > 2000}$ & $\mathbf{p_T^{\text{fatjet}} > 750}$ & \textbf{LMD $\mathbf{< 0.05}$} \\[-2pt]
	&  & \& $\mathbf{\sqrt{\hat{s}} > 2000}$ & \& $\mathbf{p_T^{\text{iso}\:\ell} > 500}$ & \& \textbf{LSF $\mathbf{> 0.9}$} \\
	\hline
	~$\mathbf{BKG\ jj}$      & 340 $\pm$ 18.44  & 55 $\pm$ 7.42  & 51 $\pm$ 7.14  & $<$ 1 \\
	~$\mathbf{BKG\ Vj}$      & 467022 $\pm$ 835.47  & 6908 $\pm$ 101.55  & 5784 $\pm$ 92.94  & 7 $\pm$ 2.56 \\
	~$\mathbf{BKG\ t\bar t}$ & 96846 $\pm$ 246.03  & 494 $\pm$ 17.58  & 417 $\pm$ 16.12  & 37 $\pm$ 4.80 \\
	~$\mathbf{BKG\ VVj}$      & 17733 $\pm$ 139.66  & 384 $\pm$ 20.56  & 327 $\pm$ 18.95  & 16 $\pm$ 4.12 \\
	\hline
	~\textbf{BP1-LH}        & 138 $\pm$ 11.75  & 75 $\pm$ 8.66  & 73 $\pm$ 8.54  & 56 $\pm$ 7.48 \\
	~\textbf{BP1-RH}        & 131 $\pm$ 11.44  & 72 $\pm$ 8.49  & 70 $\pm$ 8.37  & 54 $\pm$ 7.35 \\
	~\textbf{BP2-LH}        & 85 $\pm$ 9.22  & 51 $\pm$ 7.14  & 48 $\pm$ 6.93  & 39 $\pm$ 6.24 \\
	~\textbf{BP2-RH}        & 87 $\pm$ 9.33  & 54 $\pm$ 7.35  & 52 $\pm$ 7.21  & 40 $\pm$ 6.32 \\
	~\textbf{BP3-LH}        & 95 $\pm$ 9.75  & 73 $\pm$ 8.54  & 69 $\pm$ 8.31  & 50 $\pm$ 7.07 \\
	~\textbf{BP3-RH}        & 92 $\pm$ 9.59  & 71 $\pm$ 8.43  & 68 $\pm$ 8.25  & 50 $\pm$ 7.07 \\
	~\textbf{BP4-LH}        & 72 $\pm$ 8.49  & 56 $\pm$ 7.48  & 52 $\pm$ 7.21  & 41 $\pm$ 6.40 \\
	~\textbf{BP4-RH}        & 76 $\pm$ 8.71  & 62 $\pm$ 7.87  & 59 $\pm$ 7.68  & 46 $\pm$ 6.78 \\
	~\textbf{BP5-LH}        & 44 $\pm$ 6.63  & 38 $\pm$ 6.16  & 36 $\pm$ 6.00  & 25 $\pm$ 5.00 \\
	~\textbf{BP5-RH}        & 44 $\pm$ 6.63  & 38 $\pm$ 6.16  & 36 $\pm$ 6.00  & 25 $\pm$ 5.00 \\
	~\textbf{BP6-LH}        & 35 $\pm$ 5.92  & 31 $\pm$ 5.57  & 28 $\pm$ 5.29  & 22 $\pm$ 4.69 \\
	~\textbf{BP6-RH}        & 38 $\pm$ 6.16  & 34 $\pm$ 5.83  & 32 $\pm$ 5.66  & 25 $\pm$ 5.00 \\
	\hline
\end{tabular}
\caption{Cutflow table for background processes ($jj, \: Vj, \: t\bar{t}, \: \mathrm{and} \: VVj$) and signal processes across multiple benchmark points (BP1 -- BP6). BP$n$-LH stands for left-handed $N$, and BP$n$-RH stands for right-handed heavy neutrino in the signal for benchmark points $n = 1\; \mathrm{to} \;6$. These numbers are calculated for an integrated luminosity of $\cal L = $ 3~ab$^{-1}$ at 14 TeV LHC. Values are given as $p \pm \sqrt{p}$, where $\sqrt{p}$ represents Poisson statistical uncertainty.}
\label{tab:cutflow_witherrors}
\end{table}

The numbers of signal and background events after applying a particular cut are listed in Table~\ref{tab:cutflow_witherrors}. The `Initial' cut on the table represents the acceptance cut put at the parton-level and the requirement of an isolated lepton and a fat jet containing a lepton subjet, both having $p_T > 150$~GeV. The initial number for the dijet background is due to the requirement of isolated lepton and fat jet. In the first stage in the cut flow, we have applied basic kinematic cuts on $H_T$ and $\sqrt{\hat s}$. These two cuts have reduced the $V+$jets, $t\bar t$ and $VV+$jets to a great extent. We then applied cuts on the transverse momentum of the fat jet, $p_T^{\text{fatjet}}$, and isolated lepton, $p_T^{\text{iso}\:\ell}$. This step further eliminates a significant portion of each background.
As the cut flow progresses, we introduce cuts on the substructure variables - $LSF$ and $LMD$. These variables provide an additional layer of discrimination by identifying the unique characteristics of fat jets that contain a lepton. Specifically, the $LSF$ cut helps to isolate events where the lepton carries a substantial fraction of the fat jet's transverse momentum, while the $LMD$ cut ensures that the lepton is kinematically distinct from the rest of the jet. These substructure-based cuts are highly effective in rejecting background events where the fat jet arises from QCD processes or hadronic top decays. The combined cuts help in a significant reduction of background events while maintaining high rates of signal events at an expected integrated luminosity of $\mathcal{L} = 3$~ab$^{-1}$ (projected for HL-LHC), as shown in the corresponding Table~\ref{tab:cutflow_witherrors}.

We now compute the final signal significance ($\mathcal{S}$) anticipated at the HL-LHC, for which an integrated luminosity of $\mathcal{L} = 3$~ab$^{-1}$ is projected. The following formula is used to evaluate $\mathcal{S}$\,\cite{Cowan:2010js} for different benchmark points:
\begin{equation} 
\mathcal{S} = \sqrt{2 \left[ (S + B) \ln \left( 1 + \frac{S}{B + \epsilon^2 B (S + B)} \right) - \frac{1}{\epsilon^2} \ln \left( 1 + \frac{\epsilon^2 S}{1 + \epsilon^2 B} \right) \right]}
\end{equation}
where $S$ is the number of signal events, and $B$ is the number of background events after the full set of cuts{ , and $\epsilon$ is the overall background systematic uncertainty fraction. 
In Table~\ref{table:significance}, we show the expected significance for each benchmark point and for three selected systematic uncertainties (0\%, 5\%, and 10\%), indicating that our analysis achieves a substantial sensitivity for heavy neutrino masses in the range considered.}

\begin{table}[h!]
\centering
\begin{tabular}{|c|c|c|c|c|c|c|}
	\hline
	\multirow{2}{*}{\textbf{Benchmark Points}} & \multicolumn{2}{|c|}{\textbf{Significance ($\mathcal{S}_{\epsilon=0\%}$)}} & \multicolumn{2}{|c|}{\textbf{Significance ($\mathcal{S}_{\epsilon=5\%}$)}} & \multicolumn{2}{|c|}{\textbf{Significance ($\mathcal{S}_{\epsilon=10\%}$)}} \\
	\cline{2-7} & \qquad LH \qquad\qquad & \qquad RH\qquad\qquad & \qquad LH \qquad\qquad & \qquad RH\qquad\qquad & \qquad LH \qquad\qquad & \qquad RH\qquad\qquad\\
	\hline
	BP1 & 6.4 & 6.1 & 5.9 & 5.6 & 4.8 & 4.6 \\
	BP2 & 4.5 & 4.7 & 4.2 & 4.3 & 3.5 & 3.6 \\
	BP3 & 5.8 & 5.8 & 5.3 & 5.3 & 4.4 & 4.4 \\
	BP4 & 4.9 & 5.4 & 4.5 & 4.9 & 3.7 & 4.1 \\
	BP5 & 3.1 & 3.1 & 2.8 & 2.9 & 2.4 & 2.4 \\
	BP6 & 2.7 & 3.1 & 2.5 & 2.8 & 2.1 & 2.4 \\
	\hline
\end{tabular}
\caption{Signal significance $\mathcal{S}$ for different benchmark points. LH and RH indicate left-handed and right-handed $N$ polarizations, respectively, for each benchmark point (BP). { The values are shown for systematic uncertainties $\epsilon=0\%$, $\epsilon=5\%$, and $\epsilon=10\%$.}}
\label{table:significance}
\end{table}

From Table~\ref{table:significance}, we see the significance values for LH and RH configurations are quite close for each benchmark point, suggesting that both polarization states yield similar sensitivity in terms of event selection and background suppression, with only minor variations in significance. This is because we applied cuts on the basic kinematic variables and did not use any polarization-sensitive variable to further suppress the background. This cut set helps us first discover the heavy mass signal in the prescribed channel. One can then carry out the polarization-sensitive study, which is further discussed in Section~\ref{PolClassify}.

Across the benchmark points, certain trends emerge. {  In the absence of any systematic uncertainty, the benchmark points BP1, BP2, BP3, and BP4 exhibit significantly higher 5$\sigma$ to 6$\sigma$ signal significance for both LH and RH configurations. They remain 4$\sigma$ to 5$\sigma$ significance even after considering 10\% systematic uncertainties}, implying that these benchmarks are especially favourable for isolating the heavy neutrino signature. The enhanced significance is because of the high cross section in the low mass of $W'$. 
Conversely, benchmark points BP5 and BP6 show notably lower significance values (around 2$\sigma$ to 3$\sigma$) for both LH and RH, as the cross section is not significant for $M_{W'} = 6$~TeV.

\vspace{-6mm}
\subsection{Discrimination of LH versus RH configurations}\vspace{-3mm}
\label{PolClassify}
Once the signal region is defined, we now attempt to distinguish the LH configuration from the RH one and {\it vice versa}. For our case, the signal region (SR) has been given in Table~\ref{tab:cutflow_witherrors}, where we only used the simple kinematic event variables, without imposing any cuts on the polarization-sensitive variables. Therefore, the chosen SR is general and does not favour any particular handedness of the $W'$ couplings to the fermion pairs. We now use the polarization-sensitive variables, which we introduced in Section~\ref{sec:signal}, {\it viz.} $z_k$, $z_\ell$, and $z_\theta$, to try and distinguish the LH signal hypothesis from the RH signal hypothesis. In other words, we will set a significance score of excluding the RH structure from the LH structure (and {\it vice versa}) if such a signal is observed in the signal region at the HL-LHC. 

Our attempt of excluding one model in favour of a different model is based on $CL_s$-type method\,\cite{Feldman:1997qc,Read:2002hq}, which is essentially a profile likelihood estimator method\,\cite{Cranmer:2007zz,Cowan:2010js}. Say, for a given variable, the histogrammed distribution for two different hypotheses $H_1$ and $H_2$ has the expected number as \{$n_1$\} and \{$n_2$\}, where \{$n_i$\} is an $n$-tuple of numbers representing the bin-contents of the variable histogram corresponding to the $i^{\rm th}$ hypothesis. The confidence level of excluding $H_1$ in favour of $H_2$ is given by
\begin{equation}
CL_{H_1|H_2} = \sqrt{-2\ln \left[\frac{L\left(\{n_1\} | \{n_2\}\right)}{L\left(\{n_2\} | \{n_2\}\right)}\right]},
\end{equation}
where $L\left(\{x\} | \{y\}\right)$ represents the likelihood function. In the number counting experiments or histogrammed distribution, the likelihood is constructed assuming the Poisson distribution corresponding to each bin. Therefore, in our case, we used the likelihood function to be
\begin{equation}
L\left(\{x\} | \{y\}\right) = \prod_{k=1}^{N} \frac{e^{-x_k} \left({x_k}\right)^{y_k}}{\Gamma\left({y_k}+1\right)}, \label{eqn:CL1}
\end{equation}
where $k=1\cdots N$ runs over all the $N$ bins and $\Gamma(y+1)$ is Bernoulli's Gamma function. The interpretation of $CL_{H_1|H_2}$ is essentially the same as the significance, that is, the confidence level of rejecting hypothesis $H_1$ in favour of $H_2$.

Since one cannot really avoid the presence of background even in the best-isolated signal region, we consider the two hypotheses, in each case, to be the signal plus background. So, for the $CL_s$-type method, the hypotheses LH and RH are essentially
\begin{equation}
CL_{\rm LH|RH} = \sqrt{-2\ln \left[\frac{L\left(\{B+S_{\rm LH}\} | \{B+S_{\rm RH}\}\right)}{L\left(\{B+S_{\rm RH}\} | \{B+S_{\rm RH}\}\right)}\right]}, \label{eqn:CL2}
\end{equation} 
where $B+S_{\rm LH(RH)}$ represents the histogram of the combined background plus the LH (RH) signal of the given variable. One now chooses one or more of the three polarization-sensitive variables, {\it viz.} $z_k$, $z_\ell$, or $z_\theta$, to compute the confidence level of exclusion. In order to maximize, one may use more than one variable and create a multi-dimensional histogram to find the C.L. Due to the small number of signal and background events, the multi-dimensional distribution is dominated by statistical fluctuations and, therefore, yields an unreliable number for the discrimination power. On the other hand, a single variable cannot exploit the full potential of the combination of the three variables. We, therefore, combined the three variables by using a multivariate method, Boosted Decision Trees (BDTs), precisely XGBoost (Extreme Gradient Boosting) implementations of BDTs, to obtain a BDT classifier score as a combined variable.

\begin{figure}[!h]
\centering
\includegraphics[width=0.9\linewidth]{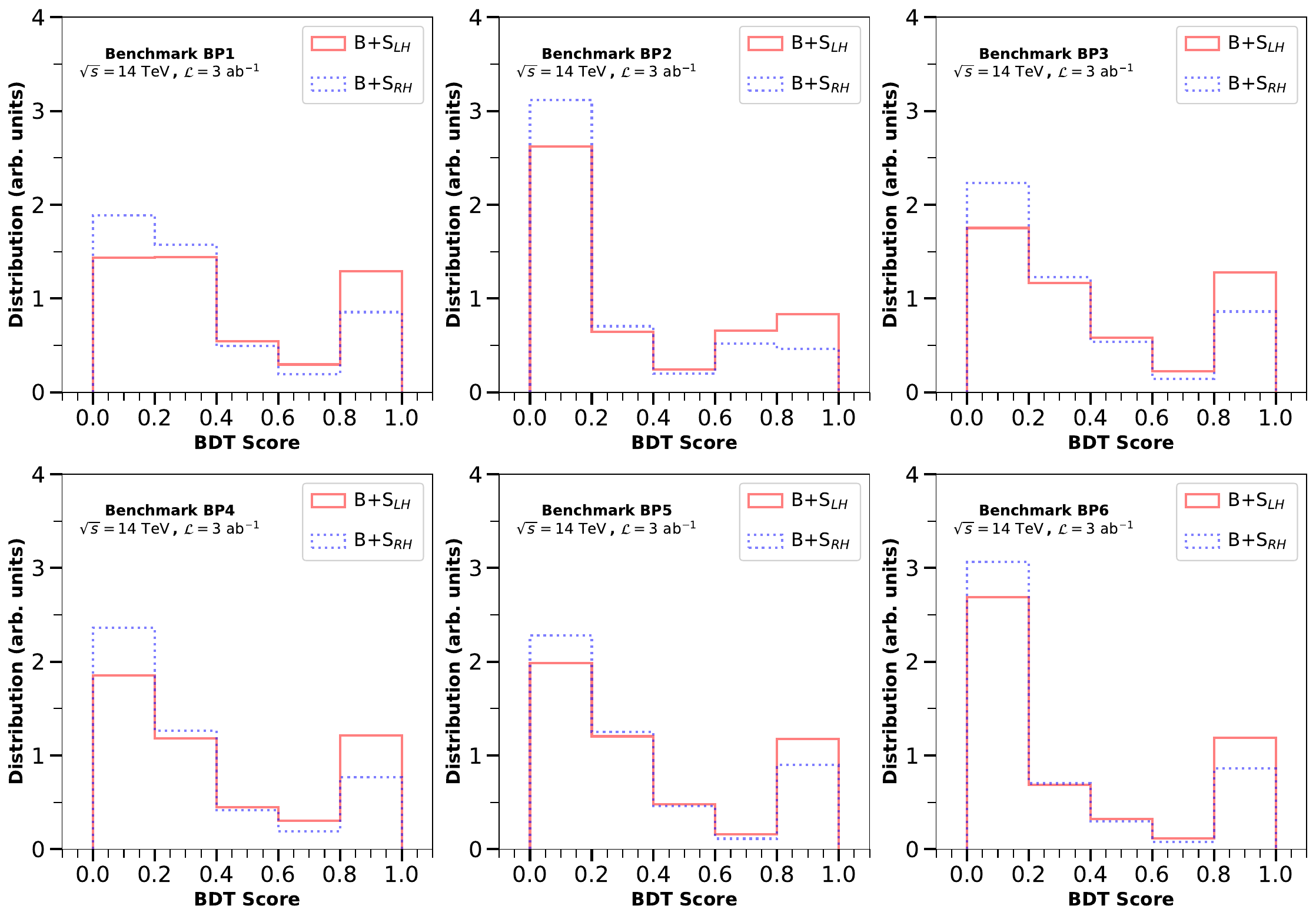}
\caption{Comparison of XGB score distributions for left-handed (LH) and right-handed (RH) signal benchmarks (BP1 to BP6) combined with background. Each subplot corresponds to a specific benchmark point to calculate $CL_{RH|LH}$.}
\label{fig:CL_LH}\vspace{-3mm}
\end{figure}

{ 
Among various implementations of BDTs, {\tt XGBoost}\,\cite{Chen_2016} is one of the most efficient and widely adopted frameworks due to its speed, scalability, and ability to handle large and complex datasets. It incorporates gradient boosting, where trees are optimized based on the gradient of the loss function, allowing for more precise learning. One of the reasons XGBoost is so effective in high-energy physics (HEP) applications is its ability to fine-tune hyperparameters to optimize performance. We, therefore, have used {\tt XGBoost}\,\cite{Chen_2016} to train the classification model and used the BDT score obtained from the classification model as the new combined variable. This BDT classifier is trained over the three polarization sensitive variables, {\it viz.} $z_k$, $z_\ell$, and $z_\theta$, of LH and RH models, excluding the SM background, as two different classes. Training has been performed for LH against RH model having the same values for $M_{W'}$ and $M_N$. Therefore a total of six  BDT models have been trained corresponding to the benchmark points BP1 to BP6 as given in Table~\ref{BP_Selection}. In each BDT training, the hyper-parameters for the classifier have been optimized to obtain a good separation score and further checked for training and testing matches to reduce overtraining. A list of important hyperparameters and ranges (across the 6 BPs) of their optimized values has been given in Table~\ref{table:param_bounds}.

\begin{table}[t]
	\centering
	\begin{tabular}{|c|p{0.7\textwidth}|c|}
		\hline
		\multirow{2}{*}{\textbf{Hyperparameter}} & \centering \multirow{2}{*}{\textbf{Description}} & \textbf{Optimal} \\
		& & \textbf{Range}\\
		\hline
		\texttt{learning\_rate} & Controls the contribution of each tree; lower values make learning slower but more robust. & [0.02, 0.08]\\
		\texttt{max\_depth} & Defines the complexity of individual decision trees. Deeper trees capture more intricate patterns but may lead to overfitting. & [4, 8] \\
		\texttt{n\_estimators} & Controls the number of boosting iterations. A higher value can improve accuracy but may lead to overfitting. & [850, 1300] \\
		\texttt{gamma} & A regularization parameter that prevents unnecessary tree splits by requiring a minimum reduction in loss. & [0.7, 3.4] \\
		\texttt{subsample} & Fraction of training data used for each boosting round to prevent overfitting. & [0.50, 0.65] \\
		\texttt{colsample\_bytree} & Determines the fraction of features used per tree, helping to reduce redundancy and improve efficiency. & [0.68, 0.95] \\
		\texttt{min\_child\_weight} & Minimum sum of instance weights needed in a child node to allow a split. & [1.0, 8.0] \\
		\texttt{lambda} & L2 regularization term to penalize large weights and reduce overfitting. & [1.2, 9.0] \\
		\texttt{alpha} & L1 regularization term to encourage sparsity in the model. & [0.2, 8.9] \\
		\texttt{scale\_pos\_weight} & Balances positive and negative classes, useful for imbalanced datasets. & [0.5, 1.1] \\
		\hline
	\end{tabular}
	\caption{  List of important hyperparameters as used by {\tt XGBoost}\,\cite{Chen_2016} with brief description of each. For each hyperparameter, the optimal range across the BPs is also listed in the last column.}
	\label{table:param_bounds}
\end{table}

\begin{table}[t]
	\centering
	\begin{tabular}{|c|c|c|c|}
		\hline
		\multirow{2}{*}{Benchmark Point} & \multicolumn{3}{c|}{Variable Importances} \\
		\cline{2-4} & \quad $z_k$~~~~ & \quad $z_\ell$ ~~~~ & \quad $z_\theta$~~~~ \\
		\hline BP1  & 0.11 & 0.58 & 0.31 \\
		\hline BP2  & 0.09 & 0.50 & 0.42 \\
		\hline BP3  & 0.09 & 0.58 & 0.32 \\
		\hline BP4  & 0.15 & 0.52 & 0.33 \\
		\hline BP5  & 0.08 & 0.58 & 0.33 \\
		\hline BP6  & 0.17 & 0.50 & 0.33 \\
		\hline
	\end{tabular}
	\caption{  Importances of the three polarization sensitive variables $z_k$, $z_\ell$, and $z_\theta$ for different benchmark points.}
	\label{tab:feat_imp}\vspace{-2mm}
\end{table}

The role of three polarization sensitive variables could be different in the discrimination between the LH and RH hypotheses and {\it vice versa}. We show their importance in the BDT training in Table~\ref{tab:feat_imp}. Overall, we can see that the variable $z_\ell$ carries the highest importance ($\gtrsim$ 50\%) followed by $z_\theta$ ($\sim$ 33\%). The lowest importance variable is $z_k$, which carries approximately 10\% importance. This is expected as $z_\ell$ distribution (as shown in Fig.~\ref{fig:partonDist_BP5}) has the least overlap between LH and RH configurations. The other two, namely $z_\theta$ and $z_k$ has more overlaps than $z_\ell$. We now show the distribution of the BDT score of signal plus SM background in the signal region in Figs.~\ref{fig:CL_LH}.}

\begin{table}[!h]
\begin{tabular}{|c|c|c|c|c|c|c|c|c|}
	\hline
	\multirow{2}{*}{Benchmark Point} & \multicolumn{4}{c|}{$CL_{\rm LH|RH}$} & \multicolumn{4}{c|}{$CL_{\rm RH|LH}$} \\
	\cline{2-9} & \quad $z_k$~~~~ & \quad $z_\ell$ ~~~~ & \quad $z_\theta$~~~~ & BDT Score & \quad $z_k$~~~~ & \quad $z_\ell$ ~~~~ & \quad $z_\theta$~~~~ & BDT Score \\
	\hline BP1  & 1.3 & 0.9 & 0.9 & 2.2 & 1.1 & 0.8 & 0.9 & 2.2 \\
	\hline BP2  & 1.6 & 1.3 & 0.9 & 1.9 & 1.3 & 1.3 & 1.0 & 1.9 \\
	\hline BP3  & 1.0 & 0.7 & 0.7 & 2.6 & 0.9 & 0.7 & 0.9 & 2.8 \\
	\hline BP4  & 2.0 & 1.6 & 1.8 & 2.4 & 0.7 & 0.8 & 1.1 & 1.6 \\
	\hline BP5  & 0.7 & 0.4 & 0.5 & 1.6 & 0.5 & 0.4 & 0.5 & 1.6 \\
	\hline BP6  & 1.2 & 0.7 & 0.7 & 1.8 & 1.0 & 0.7 & 0.7 & 1.9 \\
	\hline
\end{tabular}
\caption{Confidence level ($CL_{H_1|H_2}$) of rejecting hypothesis $H_1$ in favour of $H_2$ for three polarization-sensitive variables and the combined BDT score.}
\label{tab:CL}
\end{table}

The calculated confidence levels (according to Eq.~(\ref{eqn:CL2})) have been tabulated in Table~\ref{tab:CL}. For the columns with $z_k$, $z_\ell$ and $z_\theta$ represent the exclusion score when these variables have been used standalone. These variables could achieve the discrimination significance around 1$\sigma$ to 2$\sigma$. The columns labelled as `BDT score' represent the discrimination score when the combined variable BDT classifier score has been used. In each case, this combined variable significantly improves the discrimination score up to 2.8$\sigma$. We further note that the scores $CL_{\rm RH|LH}$ and $CL_{\rm LH|RH}$ remain similar for all variations across the benchmark points and variables, showcasing the robustness of the method.
This showcases that the jet substructure method is not only useful in finding signals by suppressing SM background, but it is also useful in constructing variables that help discriminate models with different polarization. 

\vspace{-4mm}
\section{Summary}\vspace{-3mm}
\label{sec:summary}
Many new physics scenarios predict the existence of massive charged gauge boson $W'$ and its coupling to leptons and heavy neutrinos (N). In some scenarios, this coupling appears as $V+A$ (RH) and, in others, it can be considered to be $V-A$ (LH).
The production of $W'$ and its decay to heavy neutrino and a lepton has been looked at the LHC in the $\ell\ell j j$ channel. In this work, we studied a different signal topology consisting of an energetic isolated lepton plus a fat jet, originating from the heavy $N$. This fat jet essentially embeds a lepton as a subjet along with two hadronic subjets inside it. By carefully analyzing the subjets through jet substructure-based methods and using appropriate variables, we have been able to isolate the signal by suppressing SM backgrounds. The variables $LSF$ (Lepton Subjet Fraction) and $LMD$ (Lepton Mass Drop) from the lepton subjet are particularly useful in suppressing huge QCD and $t\bar t$ backgrounds to achieve high signal significance. For the chosen benchmark points (see Table~\ref{Tab:benchmark}) with $W'$ mass ranging from 3 to 6 TeV and heavy neutrino mass ranging between 200 and 400 GeV, we have obtained $\sim 3\sigma$ to $\sim 6\sigma$ significance at the 14 TeV HL-LHC, considering 3~ab$^{-1}$ of integrated luminosity and zero systematic uncertainty. For the same benchmark points a considerable significance around $\sim 2\sigma$ to $\sim 5\sigma$ can be achieved even with 10\% systematics.

We further looked into the discrimination power between LH and RH coupling of $W'$ to lepton and heavy neutrino. We analyze this through the decay products of $N$. Since the decayed lepton from $N$ has a high spin analyzing power, appropriated variables constructed out of the lepton subjet, along with other subjets, provide a good discrimination power between the LH and RH configurations. We have used customized jet substructure-based polarization sensitive variables $z_\ell$, $z_\theta$ and $z_k$ to form the BDT score as a combined variable to help improve separation power. The discrimination score is presented in terms of confidence level by using the $CL_s$-type profile likelihood estimator method, in which the significance of excluding one hypothesis in favour of the other is obtained.
For the chosen benchmark points, we were able to achieve 1.6$\sigma$ to 2.8$\sigma$ significance of excluding the LH hypothesis in favour of RH configuration and {\it vice versa} in the signal region. Thus the 14 TeV HL-LHC with 3~ab$^{-1}$ integrated luminosity not only appears to be a good platform to probe the production of $W'$ in the lepton plus fat jet channel, it also has the potential to discriminate between LH and RH nature of coupling if such a discovery is achieved.  

\begin{acknowledgements}
The authors sincerely acknowledge Prof. Santosh Kumar Rai for insightful discussions during the initial phase of the work. The authors also acknowledge the HPC Cluster at the Regional Centre for Accelerator-based Particle Physics (RECAPP), Harish-Chandra Research Institute, and Nandadevi HPC facility, maintained and supported by the Institute of Mathematical Science's High-Performance Computing Center. T.~S.~acknowledges Anupam Ghosh for useful discussions. S.~D. acknowledges the financial support through the APEX project (theory) at the Institute of Physics, Bhubaneswar.
\end{acknowledgements}

\vspace{-4mm}
\appendix
\section{  Reinterpretation of CMS searches for $W'$ bosons with leptons and jets}
\label{AppendixA}
{ 
This appendix presents the result of the reinterpretation of the CMS searches\,\cite{CMS:2021dzb} for heavy $W'$ in the $pp \to W' \to N\ell \to \ell\ell j j$ at the 13 TeV LHC with an integrated luminosity of 138 fb$^{-1}$ luminosity. A publicly available and validated recasting\,\cite{Frank:2023epx,DVN/UMGIDL_2023} implemented within {\tt MadAnalysis5} framework\,\cite{Conte:2012fm,Conte:2014zja, Conte:2018vmg} has been used in this.
\begin{table}[t]
\centering
\renewcommand{\arraystretch}{1.2}
\begin{tabular}{|c|c|c|c|c|c|c|c|c|c|}
	\hline
	\multicolumn{2}{|c|}{Signal Region} & $N_{\rm Obs}$ & $N_{\rm SM}$ & BP1-LH & BP2-LH & BP3-LH & BP4-LH & BP5-LH & BP6-LH \\
	\hline
	\multirow{9}{*}{\rotatebox{90}{Electron Channel}}
	& (800,1000) & $1106.0 \pm 33.3$ & $1103.5 \pm 26.6$ & $3.0\!\times\! 10^{-4}$ & $1.7\!\times\! 10^{-3}$ & $1.9\!\times\! 10^{-5}$ & $1.9\!\times\! 10^{-4}$ & $0.0\!\times\! 10^{0}$ & $1.9\!\times\! 10^{-5}$ \\
	& (1000,1200) & $646.0 \pm 25.4$ & $631.5 \pm 17.0$ & $8.9\!\times\! 10^{-4}$ & $4.9\!\times\! 10^{-3}$ & $3.9\!\times\! 10^{-5}$ & $5.6\!\times\! 10^{-4}$ & $2.9\!\times\! 10^{-5}$ & $8.7\!\times\! 10^{-5}$ \\
	& (1200,1400) & $332.0 \pm 18.2$ & $323.2 \pm 10.7$ & $1.6\!\times\! 10^{-3}$ & $5.2\!\times\! 10^{-3}$ & $3.3\!\times\! 10^{-4}$ & $1.4\!\times\! 10^{-3}$ & $0.0\!\times\! 10^{0}$ & $9.7\!\times\! 10^{-5}$ \\
	& (1400,1600) & $170.0 \pm 13.0$ & $169.7 \pm 6.8$ & $2.1\!\times\! 10^{-3}$ & $5.8\!\times\! 10^{-3}$ & $1.9\!\times\! 10^{-4}$ & $2.0\!\times\! 10^{-3}$ & $1.9\!\times\! 10^{-5}$ & $1.6\!\times\! 10^{-4}$ \\
	& (1600,2000) & $143.0 \pm 12.0$ & $157.6 \pm 9.5$ & $5.2\!\times\! 10^{-3}$ & $1.6\!\times\! 10^{-2}$ & $1.0\!\times\! 10^{-3}$ & $4.1\!\times\! 10^{-3}$ & $1.7\!\times\! 10^{-4}$ & $5.4\!\times\! 10^{-4}$ \\
	& (2000,2400) & $62.0 \pm 7.9$ & $52.3 \pm 4.0$ & $6.5\!\times\! 10^{-3}$ & $1.8\!\times\! 10^{-2}$ & $1.3\!\times\! 10^{-3}$ & $4.7\!\times\! 10^{-3}$ & $1.9\!\times\! 10^{-4}$ & $8.8\!\times\! 10^{-4}$ \\
	& (2400,2800) & $25.0 \pm 5.0$ & $19.6 \pm 1.5$ & $7.4\!\times\! 10^{-3}$ & $2.2\!\times\! 10^{-2}$ & $9.3\!\times\! 10^{-4}$ & $5.8\!\times\! 10^{-3}$ & $1.4\!\times\! 10^{-4}$ & $1.1\!\times\! 10^{-3}$ \\
	& (2800,3200) & $10.0 \pm 3.2$ & $9.0 \pm 1.2$ & $1.0\!\times\! 10^{-2}$ & $3.3\!\times\! 10^{-2}$ & $1.3\!\times\! 10^{-3}$ & $7.4\!\times\! 10^{-3}$ & $1.9\!\times\! 10^{-4}$ & $1.3\!\times\! 10^{-3}$ \\
	& (3200,8000) & $13.0 \pm 3.6$ & $6.2 \pm 0.8$ & $1.3\!\times\! 10^{-1}$ & $4.4\!\times\! 10^{-1}$ & $4.0\!\times\! 10^{-2}$ & $2.3\!\times\! 10^{-1}$ & $4.6\!\times\! 10^{-3}$ & $3.0\!\times\! 10^{-2}$ \\
	\hline
	\multirow{9}{*}{\rotatebox{90}{Muon Channel}}
	& (800,1000) & $1639.0 \pm 40.5$ & $1670.7 \pm 39.8$ & $1.9\!\times\! 10^{-3}$ & $4.3\!\times\! 10^{-3}$ & $9.7\!\times\! 10^{-5}$ & $6.6\!\times\! 10^{-4}$ & $0.0\!\times\! 10^{0}$ & $5.8\!\times\! 10^{-5}$ \\
	& (1000,1200) & $946.0 \pm 30.8$ & $926.0 \pm 23.9$ & $3.2\!\times\! 10^{-3}$ & $7.2\!\times\! 10^{-3}$ & $2.1\!\times\! 10^{-4}$ & $1.3\!\times\! 10^{-3}$ & $1.9\!\times\! 10^{-5}$ & $1.4\!\times\! 10^{-4}$ \\
	& (1200,1400) & $518.0 \pm 22.8$ & $500.3 \pm 14.9$ & $4.3\!\times\! 10^{-3}$ & $1.2\!\times\! 10^{-2}$ & $3.7\!\times\! 10^{-4}$ & $2.2\!\times\! 10^{-3}$ & $3.9\!\times\! 10^{-5}$ & $3.2\!\times\! 10^{-4}$ \\
	& (1400,1600) & $268.0 \pm 16.4$ & $263.9 \pm 9.3$ & $6.0\!\times\! 10^{-3}$ & $1.6\!\times\! 10^{-2}$ & $9.3\!\times\! 10^{-4}$ & $3.5\!\times\! 10^{-3}$ & $1.2\!\times\! 10^{-4}$ & $6.3\!\times\! 10^{-4}$ \\
	& (1600,2000) & $216.0 \pm 14.7$ & $215.2 \pm 8.2$ & $1.8\!\times\! 10^{-2}$ & $4.6\!\times\! 10^{-2}$ & $2.1\!\times\! 10^{-3}$ & $9.6\!\times\! 10^{-3}$ & $2.3\!\times\! 10^{-4}$ & $1.6\!\times\! 10^{-3}$ \\
	& (2000,2400) & $80.0 \pm 8.9$ & $73.5 \pm 4.5$ & $2.4\!\times\! 10^{-2}$ & $6.4\!\times\! 10^{-2}$ & $3.1\!\times\! 10^{-3}$ & $1.4\!\times\! 10^{-2}$ & $4.3\!\times\! 10^{-4}$ & $2.4\!\times\! 10^{-3}$ \\
	& (2400,2800) & $30.0 \pm 5.5$ & $25.9 \pm 2.3$ & $3.7\!\times\! 10^{-2}$ & $9.0\!\times\! 10^{-2}$ & $4.3\!\times\! 10^{-3}$ & $1.9\!\times\! 10^{-2}$ & $5.2\!\times\! 10^{-4}$ & $3.6\!\times\! 10^{-3}$ \\
	& (2800,3200) & $13.0 \pm 3.6$ & $9.8 \pm 1.2$ & $2.5\!\times\! 10^{-1}$ & $5.2\!\times\! 10^{-1}$ & $5.4\!\times\! 10^{-3}$ & $2.4\!\times\! 10^{-2}$ & $6.5\!\times\! 10^{-4}$ & $4.0\!\times\! 10^{-3}$ \\
	& (3200,8000) & $11.0 \pm 3.3$ & $7.8 \pm 0.8$ & $8.6\!\times\! 10^{-2}$ & $1.9\!\times\! 10^{-1}$ & $1.2\!\times\! 10^{-1}$ & $4.4\!\times\! 10^{-1}$ & $3.0\!\times\! 10^{-2}$ & $1.5\!\times\! 10^{-1}$ \\
	\hline
\end{tabular}
\caption{  Comparison of the number of surviving events in the signal regions as defined in CMS search\,\cite{CMS:2021dzb}. The ranges in the Signal Region represent the ranges of $\ell\ell j j$ invariant mass corresponding to electron and muon channels. The number $N_{\rm Obs}$, $N_{\rm SM}$ are as reported by CMS collaboration\,\cite{hepdata.114866}. The last six columns are for benchmark points for BP1-LH -- BP6-LH after analysis performed using the validated analysis\,\cite{DVN/UMGIDL_2023}. For each benchmark point, a total of 5 million events were generated at the {\tt MadGraph5} parton-level.}
\label{tab:recast_LH}
\end{table}
For this part, the signal events for $pp \to W' \to N\ell \to \ell\ell j j$ have been generated in the same way as described in section~\ref{sec:signal}.
The {\tt PYTHIA-8.3} hadron-level events have been taken as input to the validated analysis. The recasting analysis incorporates the detector-level effects\,\cite{DVN/UMGIDL_2023}. The number of surviving signal events in each signal region characterized by the invariant mass for $eejj$ and $\mu\mu jj$ as defined in Ref.\,\cite{CMS:2021dzb} have been tabulated in Tables~\ref{tab:recast_LH} and \ref{tab:recast_RH} for the LH and RH benchmark points. Each table also contains the number of observed events ($N_{\rm Obs}$) at CMS and the number of expected background events ($N_{\rm SM}$) for all signal regions. In all signal regions, we see the number of signal events predicted after recasting from our model is much less than the expected observed and background uncertainties. Therefore, we can safely assume all chosen Benchmark points are allowed by the analysis. The fractional numbers in the model prediction represent the scaling down of the large number of generated Monte Carlo events (5M for each BP) to the predicted for an integrated luminosity of 138~fb$^{-1}$ of the CMS search.
\begin{table}[t]
\centering
\renewcommand{\arraystretch}{1.2}
\begin{tabular}{|c|c|c|c|c|c|c|c|c|c|}
	\hline
	\multicolumn{2}{|c|}{Signal Region}  & $N_{\rm Obs}$ & $N_{\rm SM}$ & BP1-RH & BP2-RH & BP3-RH & BP4-RH & BP5-RH & BP6-RH \\
	\hline
	\multirow{9}{*}{\rotatebox{90}{Electron Channel}}
	& (800,1000) & $1106.0 \pm 33.3$ & $1103.5 \pm 26.6$ & $5.4\!\times\! 10^{-4}$ & $2.6\!\times\! 10^{-3}$ & $1.4\!\times\! 10^{-4}$ & $2.5\!\times\! 10^{-4}$ & $9.7\!\times\! 10^{-6}$ & $7.7\!\times\! 10^{-5}$ \\
	& (1000,1200) & $646.0 \pm 25.4$ & $631.5 \pm 17.0$ & $1.6\!\times\! 10^{-3}$ & $5.4\!\times\! 10^{-3}$ & $1.4\!\times\! 10^{-4}$ & $8.5\!\times\! 10^{-4}$ & $9.7\!\times\! 10^{-6}$ & $1.2\!\times\! 10^{-4}$ \\
	& (1200,1400) & $332.0 \pm 18.2$ & $323.2 \pm 10.7$ & $2.5\!\times\! 10^{-3}$ & $6.9\!\times\! 10^{-3}$ & $2.7\!\times\! 10^{-4}$ & $2.0\!\times\! 10^{-3}$ & $3.9\!\times\! 10^{-5}$ & $1.4\!\times\! 10^{-4}$ \\
	& (1400,1600) & $170.0 \pm 13.0$ & $169.7 \pm 6.8$ & $3.1\!\times\! 10^{-3}$ & $7.6\!\times\! 10^{-3}$ & $6.4\!\times\! 10^{-4}$ & $2.4\!\times\! 10^{-3}$ & $6.8\!\times\! 10^{-5}$ & $3.2\!\times\! 10^{-4}$ \\
	& (1600,2000) & $143.0 \pm 12.0$ & $157.6 \pm 9.5$ & $6.7\!\times\! 10^{-3}$ & $1.8\!\times\! 10^{-2}$ & $1.3\!\times\! 10^{-3}$ & $6.2\!\times\! 10^{-3}$ & $2.5\!\times\! 10^{-4}$ & $1.3\!\times\! 10^{-3}$ \\
	& (2000,2400) & $62.0 \pm 7.9$ & $52.3 \pm 4.0$ & $8.2\!\times\! 10^{-3}$ & $2.3\!\times\! 10^{-2}$ & $1.7\!\times\! 10^{-3}$ & $6.3\!\times\! 10^{-3}$ & $4.4\!\times\! 10^{-4}$ & $1.4\!\times\! 10^{-3}$ \\
	& (2400,2800) & $25.0 \pm 5.0$ & $19.6 \pm 1.5$ & $9.3\!\times\! 10^{-3}$ & $2.6\!\times\! 10^{-2}$ & $2.1\!\times\! 10^{-3}$ & $7.2\!\times\! 10^{-3}$ & $2.7\!\times\! 10^{-4}$ & $1.5\!\times\! 10^{-3}$ \\
	& (2800,3200) & $10.0 \pm 3.2$ & $9.0 \pm 1.2$ & $1.5\!\times\! 10^{-2}$ & $4.0\!\times\! 10^{-2}$ & $1.9\!\times\! 10^{-3}$ & $8.4\!\times\! 10^{-3}$ & $2.9\!\times\! 10^{-4}$ & $1.8\!\times\! 10^{-3}$ \\
	& (3200,8000) & $13.0 \pm 3.6$ & $6.2 \pm 0.8$ & $1.6\!\times\! 10^{-1}$ & $4.4\!\times\! 10^{-1}$ & $6.2\!\times\! 10^{-2}$ & $2.9\!\times\! 10^{-1}$ & $8.3\!\times\! 10^{-3}$ & $4.5\!\times\! 10^{-2}$ \\
	\hline
	\multirow{9}{*}{\rotatebox{90}{Muon Channel}}
	& (800,1000) & $1639.0 \pm 40.5$ & $1670.7 \pm 39.8$ & $1.1\!\times\! 10^{-3}$ & $2.4\!\times\! 10^{-3}$ & $5.8\!\times\! 10^{-5}$ & $4.1\!\times\! 10^{-4}$ & $9.7\!\times\! 10^{-6}$ & $1.1\!\times\! 10^{-4}$ \\
	& (1000,1200) & $946.0 \pm 30.8$ & $926.0 \pm 23.9$ & $2.4\!\times\! 10^{-3}$ & $6.0\!\times\! 10^{-3}$ & $3.7\!\times\! 10^{-4}$ & $8.9\!\times\! 10^{-4}$ & $2.9\!\times\! 10^{-5}$ & $1.5\!\times\! 10^{-4}$ \\
	& (1200,1400) & $518.0 \pm 22.8$ & $500.3 \pm 14.9$ & $3.2\!\times\! 10^{-3}$ & $9.3\!\times\! 10^{-3}$ & $5.0\!\times\! 10^{-4}$ & $1.3\!\times\! 10^{-3}$ & $8.7\!\times\! 10^{-5}$ & $3.4\!\times\! 10^{-4}$ \\
	& (1400,1600) & $268.0 \pm 16.4$ & $263.9 \pm 9.3$ & $5.1\!\times\! 10^{-3}$ & $1.2\!\times\! 10^{-2}$ & $7.9\!\times\! 10^{-4}$ & $2.7\!\times\! 10^{-3}$ & $3.9\!\times\! 10^{-5}$ & $4.6\!\times\! 10^{-4}$ \\
	& (1600,2000) & $216.0 \pm 14.7$ & $215.2 \pm 8.2$ & $1.2\!\times\! 10^{-2}$ & $3.3\!\times\! 10^{-2}$ & $2.0\!\times\! 10^{-3}$ & $7.8\!\times\! 10^{-3}$ & $1.5\!\times\! 10^{-4}$ & $1.7\!\times\! 10^{-3}$ \\
	& (2000,2400) & $80.0 \pm 8.9$ & $73.5 \pm 4.5$ & $1.8\!\times\! 10^{-2}$ & $4.9\!\times\! 10^{-2}$ & $2.2\!\times\! 10^{-3}$ & $1.2\!\times\! 10^{-2}$ & $4.8\!\times\! 10^{-4}$ & $2.2\!\times\! 10^{-3}$ \\
	& (2400,2800) & $30.0 \pm 5.5$ & $25.9 \pm 2.3$ & $2.9\!\times\! 10^{-2}$ & $7.2\!\times\! 10^{-2}$ & $3.8\!\times\! 10^{-3}$ & $1.5\!\times\! 10^{-2}$ & $4.9\!\times\! 10^{-4}$ & $3.1\!\times\! 10^{-3}$ \\
	& (2800,3200) & $13.0 \pm 3.6$ & $9.8 \pm 1.2$ & $1.9\!\times\! 10^{-1}$ & $4.0\!\times\! 10^{-1}$ & $4.1\!\times\! 10^{-3}$ & $1.8\!\times\! 10^{-2}$ & $5.4\!\times\! 10^{-4}$ & $3.8\!\times\! 10^{-3}$ \\
	& (3200,8000) & $11.0 \pm 3.3$ & $7.8 \pm 0.8$ & $9.2\!\times\! 10^{-2}$ & $2.0\!\times\! 10^{-1}$ & $1.1\!\times\! 10^{-1}$ & $4.0\!\times\! 10^{-1}$ & $2.9\!\times\! 10^{-2}$ & $1.4\!\times\! 10^{-1}$ \\		\hline
\end{tabular}
\caption{  Same as Table~\ref{tab:recast_LH} but for RH cases.}
\label{tab:recast_RH}
\end{table}
}

\section{  Cross Section and Number of Simulated Events}
\label{AppendixB}
{  The cross sections and number of events generated at the {\tt MadGraph5} parton-level events are given in Table~\ref{tab:cross_section} for different signal benchmark points and backgrounds. The cross sections are provided after an `Initial cut' as described in Section~\ref{sec:bkg}.}

\begin{table}[h]
\centering
\begin{tabular}{|>{\arraybackslash}p{3cm}|>{\centering\arraybackslash}p{3cm}|c|}
	\hline
	\multirow{2}{*}{\quad\textbf{Process}} & \multirow{2}{*}{\textbf{Cross section [fb]}} & \multirow{2}{*}{\textbf{No. of Events}} \\[-2pt]
	&  &  \\
	\hline
	~$\mathbf{BKG\ jj}$      & 36602.99  & 30,000,000 \\
	~$\mathbf{BKG\ Vj}$      & 9400.7   & 21,430,000 \\
	~$\mathbf{BKG\ VVj}$     & 877.8     & 1,000,000 \\
	~$\mathbf{BKG\ t\bar t}$ & 3472.0    & 5,000,000 \\
	\hline
	~\textbf{BP1}        & 0.098      & 1,000,000 \\
	~\textbf{BP2}        & 0.097      & 1,000,000 \\
	~\textbf{BP3}        & 0.07      & 1,000,000 \\
	~\textbf{BP4}        & 0.07      & 1,000,000 \\
	~\textbf{BP5}        & 0.035     & 1,000,000 \\
	~\textbf{BP6}        & 0.035     & 1,000,000 \\
	\hline
\end{tabular}
\caption{Cross section and number of events simulated for background processes ($jj, \: V+$jets, $\: t\bar{t}, \: \mathrm{and} \: VV+$jets) and signal processes across multiple benchmark points (BP1--BP6). In each BP, the cross sections and number of simulated events remain the same for LH and RH configurations. 
}
\label{tab:cross_section}
\end{table}

\section{Distribution of variables for different benchmark points}
\label{AppendixC}\vspace{-1mm}
\subsection{Distribution of the polarization-sensitive variables}\vspace{-5mm}
\begin{figure}[!h]
\centering
\includegraphics[width=0.95\linewidth]{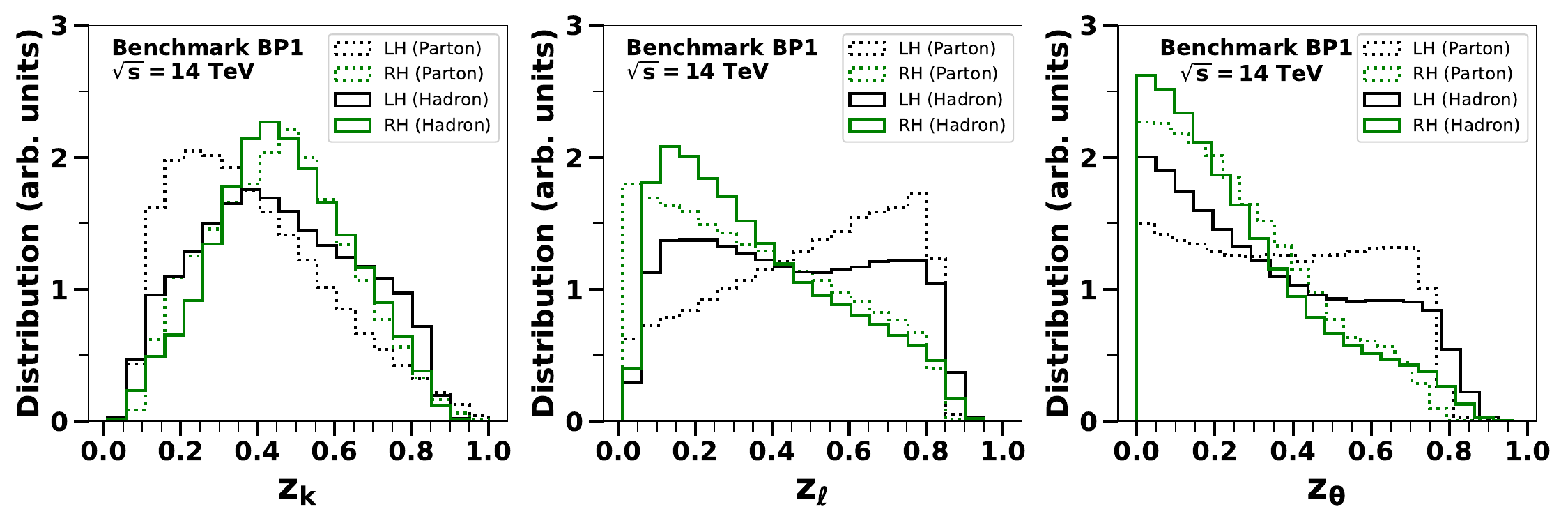}
\includegraphics[width=0.95\linewidth]{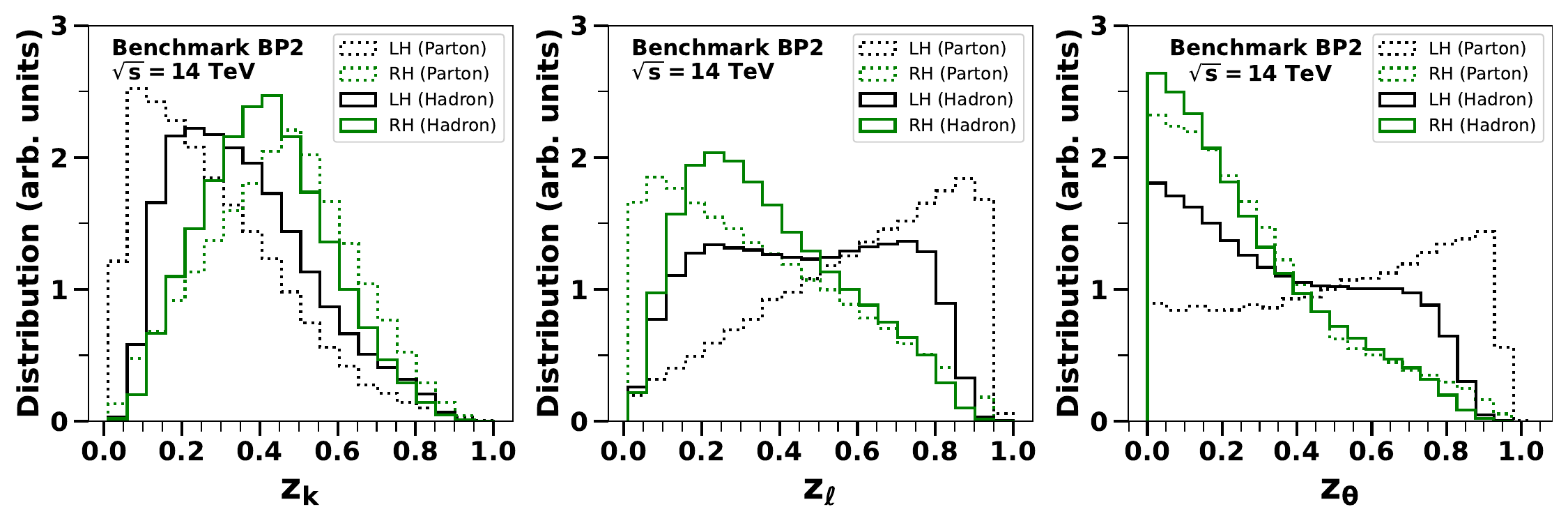}\vspace{-5mm}
\caption{Same as Fig.~\ref{fig:partonDist_BP5} but for BP1 (top row) and BP2 (bottom row).}
\label{fig:partonDist_BP1}
\end{figure}

\clearpage
\begin{figure}[h]
\centering
\includegraphics[width=0.95\linewidth]{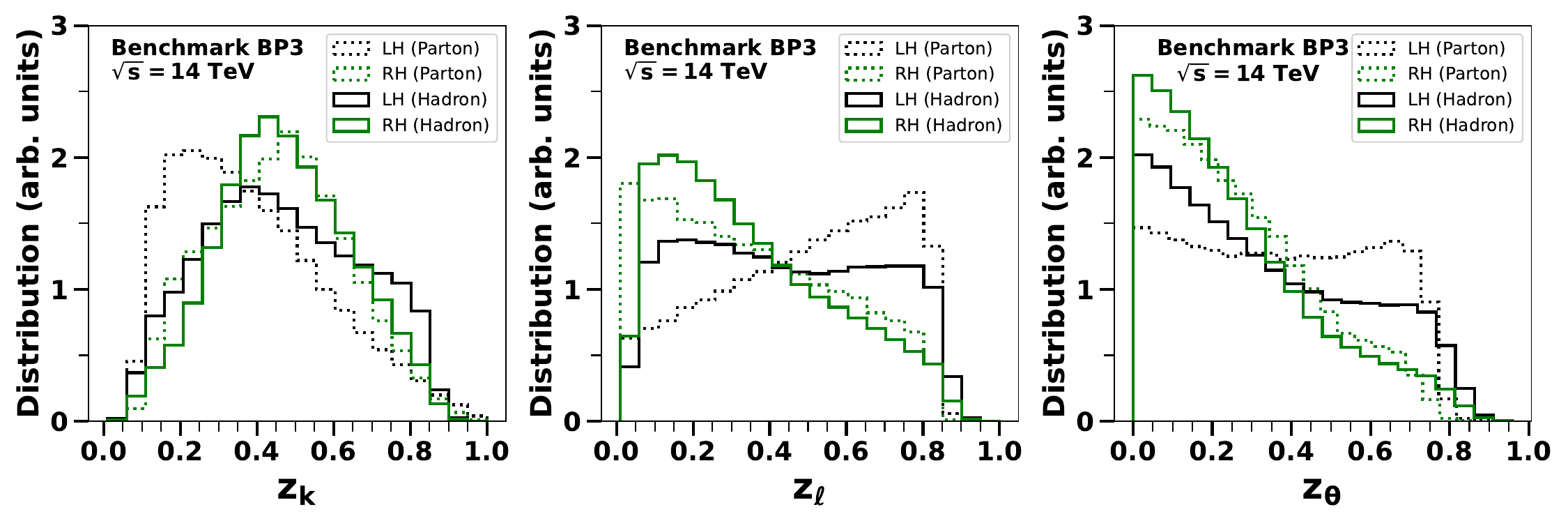}
\includegraphics[width=0.95\linewidth]{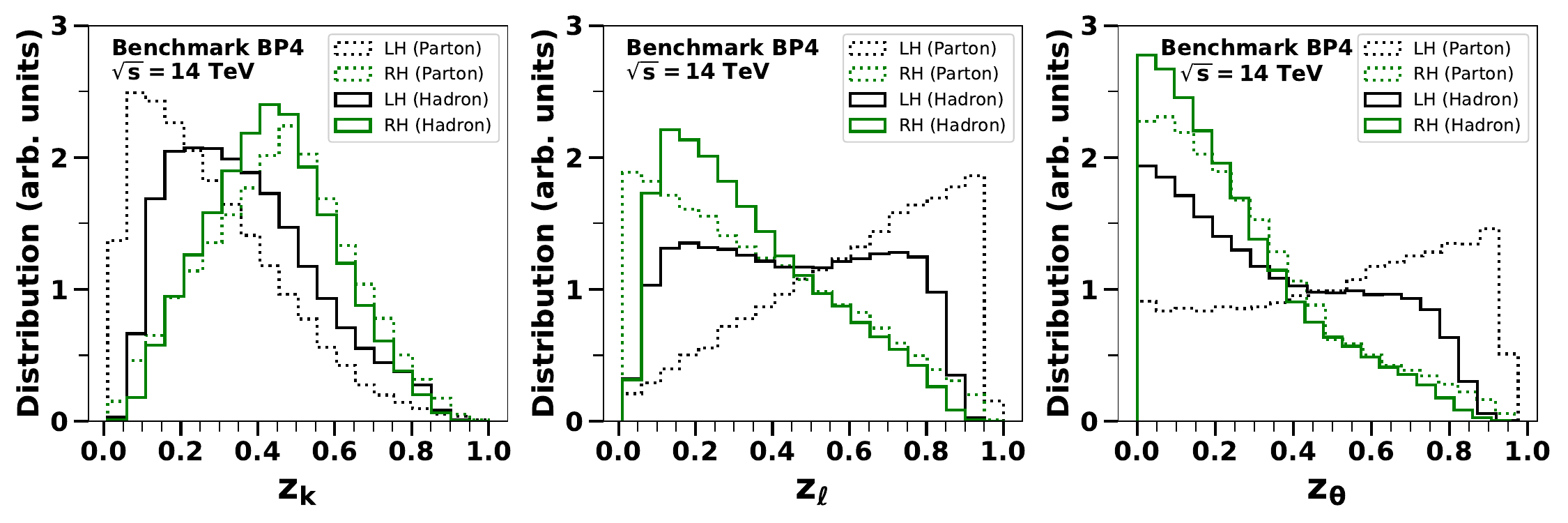}\vspace{-5mm}
\caption{Same as Fig.~\ref{fig:partonDist_BP5} but for BP3 (top row) and BP4 (bottom row).}
\label{fig:partonDist_BP3}
\end{figure}

\subsection{Distribution of key kinematics variables}
\vspace{-8mm}
\begin{figure}[h]
\centering
\includegraphics[width=\linewidth]{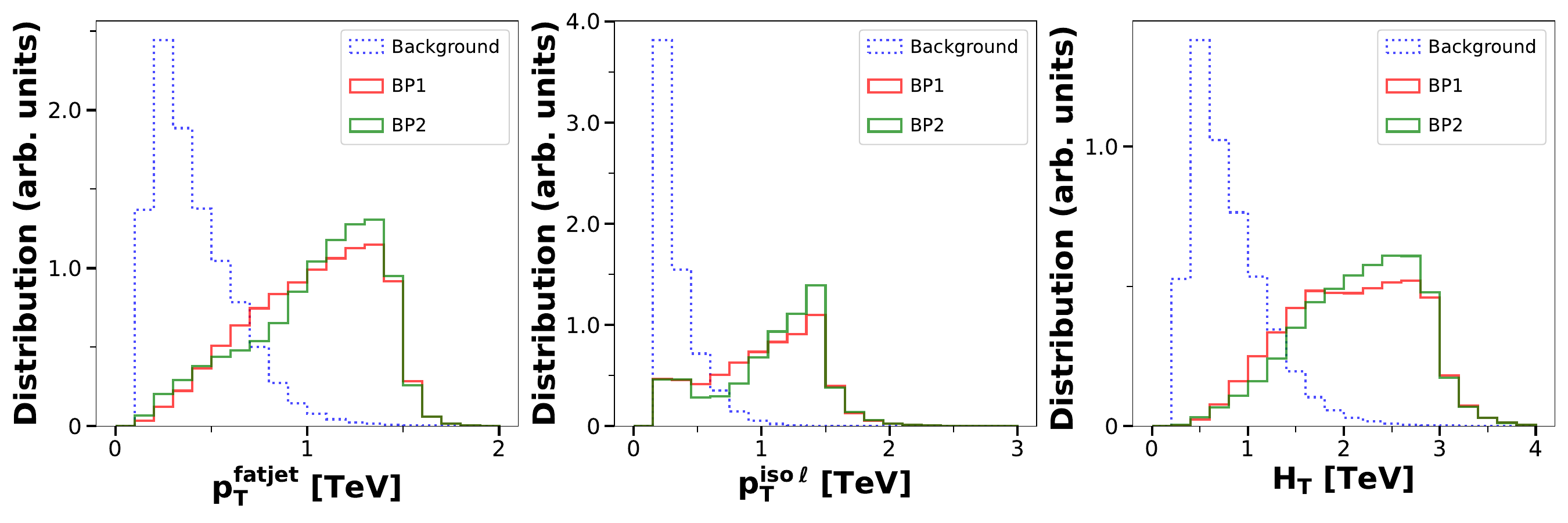}
\caption{Distributions of key kinematic variables for the signal and background processes for BP1 and BP2.}\vspace{14mm}
\label{fig:KinVar_BP12_set1}
\end{figure}

\begin{figure}[h]
\centering
\includegraphics[width=\linewidth]{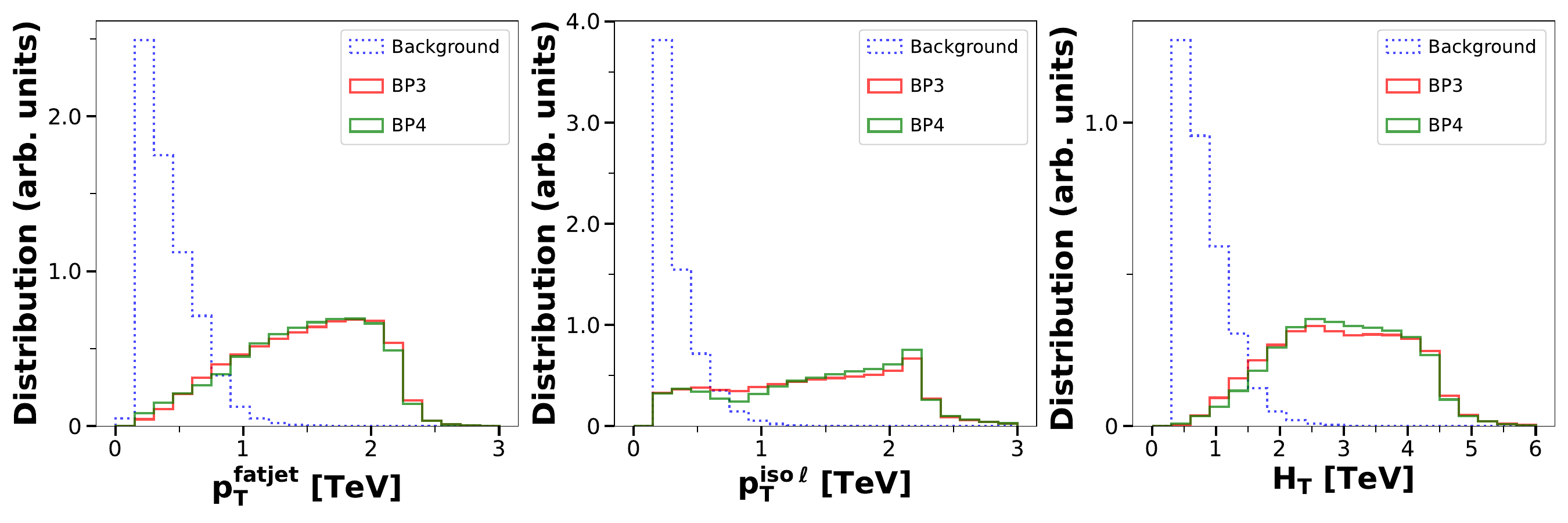}\vspace{-5mm}
\caption{Distributions of key kinematic variables for the signal and background processes for BP3 and BP4.}
\label{fig:KinVar_BP34_Set1}
\end{figure}

\subsection{Distribution of jet-substructure based variables}
\vspace{-5mm}
\begin{figure}[H]
\centering
\includegraphics[width=0.8\linewidth]{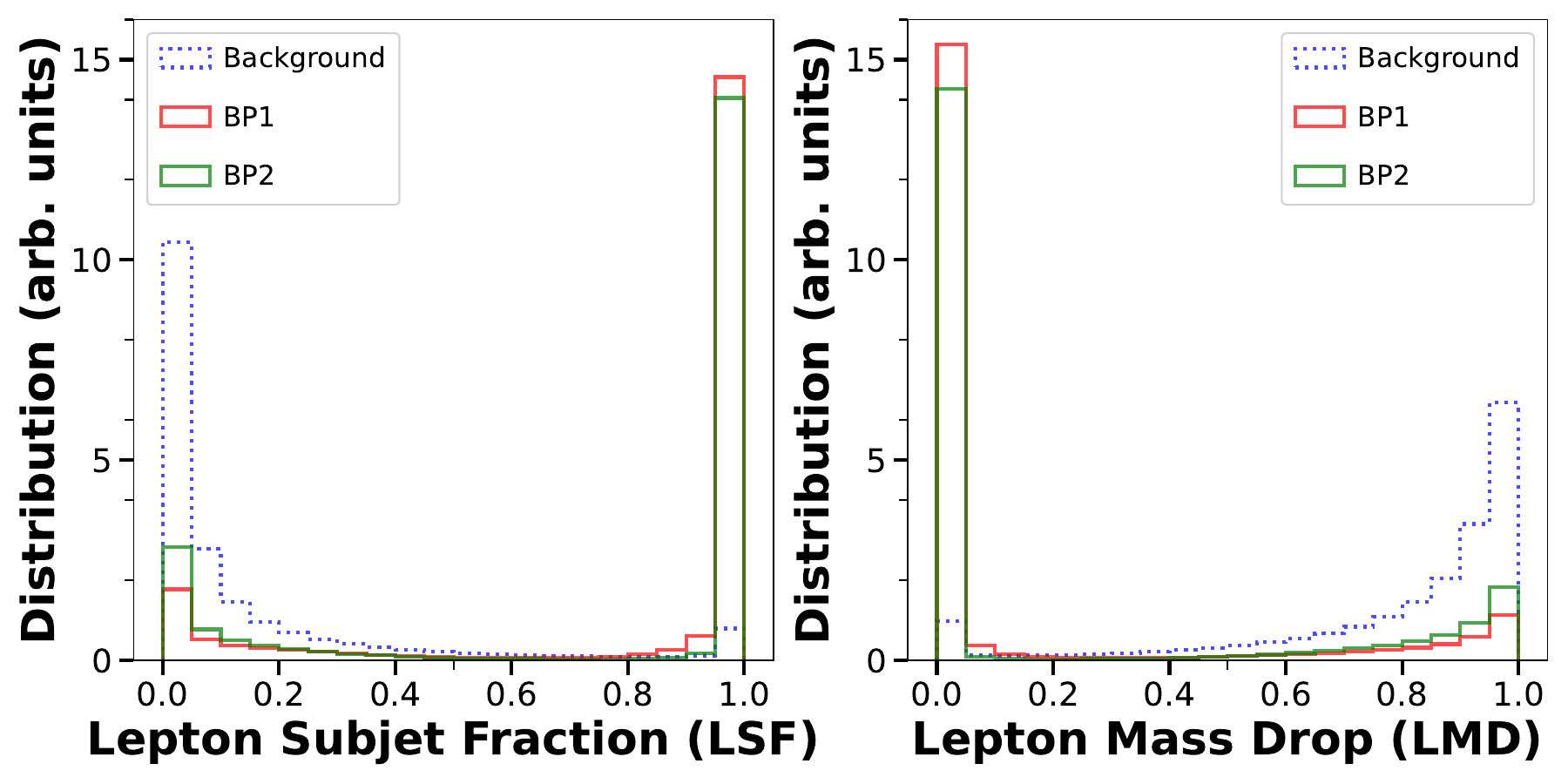}
\includegraphics[width=0.8\linewidth]{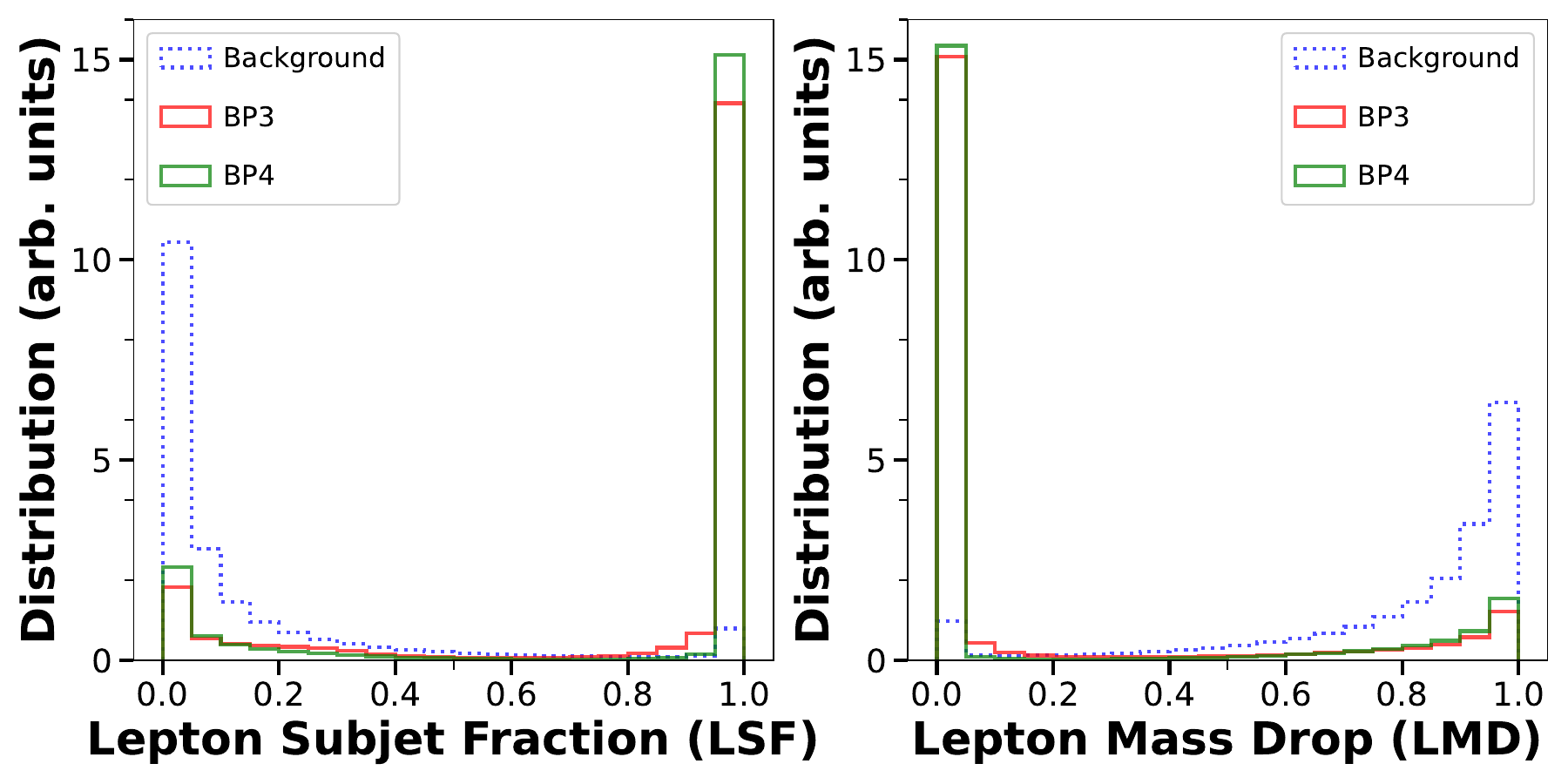}
\caption{Distributions of jet substructure variables for the signal and background processes for BP1, BP2, BP3 and BP4.}
\label{fig:KinVar_BP34_Set2}
\end{figure}

\bibliographystyle{JHEP}
\bibliography{refs}
\end{document}